\documentclass[12pt,a4paper]{article}

\usepackage{epsfig}
\usepackage{amstex}

\def\ap#1#2#3   {{\em Ann. Phys. (NY)} {\bf#1} (#2) #3}
\def\apj#1#2#3  {{\em Astrophys. J.} {\bf#1} (#2) #3}
\def\apjl#1#2#3 {{\em Astrophys. J. Lett.} {\bf#1} (#2) #3}
\def\app#1#2#3  {{\em Acta. Phys. Pol.} {\bf#1} (#2) #3}
\def\ar#1#2#3   {{\em Ann. Rev. Nucl. Part. Sci.} {\bf#1} (#2) #3}
\def\cpc#1#2#3  {{\em Computer Phys. Comm.} {\bf#1} (#2) #3}
\def\err#1#2#3  {{\it Erratum} {\bf#1} (#2) #3}
\def\ib#1#2#3   {{\it ibid.} {\bf#1} (#2) #3}
\def\jmp#1#2#3  {{\em J. Math. Phys.} {\bf#1} (#2) #3}
\def\ijmp#1#2#3 {{\em Int. J. Mod. Phys.} {\bf#1} (#2) #3}
\def\jetp#1#2#3 {{\em JETP Lett.} {\bf#1} (#2) #3}
\def\jpg#1#2#3  {{\em J. Phys. G.} {\bf#1} (#2) #3}
\def\mpl#1#2#3  {{\em Mod. Phys. Lett.} {\bf#1} (#2) #3}
\def\nat#1#2#3  {{\em Nature (London)} {\bf#1} (#2) #3}
\def\nc#1#2#3   {{\em Nuovo Cim.} {\bf#1} (#2) #3}
\def\nim#1#2#3  {{\em Nucl. Instr. Meth.} {\bf#1} (#2) #3}
\def\np#1#2#3   {{\em Nucl. Phys.} {\bf#1} (#2) #3}
\def\pcps#1#2#3 {{\em Proc. Cam. Phil. Soc.} {\bf#1} (#2) #3}
\def\pl#1#2#3   {{\em Phys. Lett.} {\bf#1} (#2) #3}
\def\prep#1#2#3 {{\em Phys. Rep.} {\bf#1} (#2) #3}
\def\prev#1#2#3 {{\em Phys. Rev.} {\bf#1} (#2) #3}
\def\prl#1#2#3  {{\em Phys. Rev. Lett.} {\bf#1} (#2) #3}
\def\prs#1#2#3  {{\em Proc. Roy. Soc.} {\bf#1} (#2) #3}
\def\ptp#1#2#3  {{\em Prog. Th. Phys.} {\bf#1} (#2) #3}
\def\ps#1#2#3   {{\em Physica Scripta} {\bf#1} (#2) #3}
\def\rmp#1#2#3  {{\em Rev. Mod. Phys.} {\bf#1} (#2) #3}
\def\rpp#1#2#3  {{\em Rep. Prog. Phys.} {\bf#1} (#2) #3}
\def\sjnp#1#2#3 {{\em Sov. J. Nucl. Phys.} {\bf#1} (#2) #3}
\def\zp#1#2#3   {{\em Z. Phys.} {\bf#1} (#2) #3}

\def\nostrocostrutto#1\over#2{\mathrel{\mathop{\kern 0pt \rlap 
  {\raise.2ex\hbox{$#1$}}}
  \lower.9ex\hbox{\kern-.190em $#2$}}}

\newcommand{\labl}[1]{\label{#1}}

\newcommand{\eref}[1]{(\ref{#1})}      

\newcommand{\kc}{k^{cut}}
\newcommand{\kct}{k_{\perp}^{cut}}
\newcommand{\kt}{k_{\perp}}
\newcommand{\ad}{{\gamma_{0}^{2}}}
\newcommand{\xt}{X_{\perp}}                  

\catcode`@=11
\newcount\@tempcntc
\def\@citex[#1]#2{\if@filesw\immediate\write\@auxout{\string\citation{#2}}\fi
  \@tempcnta\z@\@tempcntb\m@ne\def\@citea{}\@cite{\@for\@citeb:=#2\do
    {\@ifundefined
       {b@\@citeb}{\@citeo\@tempcntb\m@ne\@citea\def\@citea{,}{\bf ?}\@warning
       {Citation `\@citeb' on page \thepage \space undefined}}%
    {\setbox\z@\hbox{\global\@tempcntc0\csname b@\@citeb\endcsname\relax}%
     \ifnum\@tempcntc=\z@ \@citeo\@tempcntb\m@ne
       \@citea\def\@citea{,}\hbox{\csname b@\@citeb\endcsname}%
     \else
      \advance\@tempcntb\@ne
      \ifnum\@tempcntb=\@tempcntc
      \else\advance\@tempcntb\m@ne\@citeo
      \@tempcnta\@tempcntc\@tempcntb\@tempcntc\fi\fi}}\@citeo}{#1}}
\def\@citeo{\ifnum\@tempcnta>\@tempcntb\else\@citea\def\@citea{,}%
  \ifnum\@tempcnta=\@tempcntb\the\@tempcnta\else
   {\advance\@tempcnta\@ne\ifnum\@tempcnta=\@tempcntb \else \def\@citea{--}\fi
    \advance\@tempcnta\m@ne\the\@tempcnta\@citea\the\@tempcntb}\fi\fi}
\catcode`@=12

\begin{document}

\setcounter{page}{0}
\thispagestyle{empty}

\rightline{MPI-PhT/98-31}
\rightline{TPJU-7/98}
\rightline{April 27th, 1998}

\vfill 

\begin{center}
{\Large \bf Poissonian limit of  soft gluon multiplicity}
\end{center}
\vfill 
\begin{center}
SERGIO LUPIA~$^{a, }$\footnote{e-mail: lupia@mppmu.mpg.de} \ , \ 
WOLFGANG OCHS~$^{a, }$\footnote{e-mail: wwo@mppmu.mpg.de} \ ,\
JACEK WOSIEK~$^{b, }$\footnote{e-mail: wosiek@thrisc.if.uj.edu.pl}

\end{center}

\begin{center}
$^a$ \ {\it Max-Planck-Institut f\"ur Physik \\
(Werner-Heisenberg-Institut) \\
F\"ohringer Ring 6, D-80805 Munich, Germany} \\
\mbox{ }\\
$^b$ \ {\it Jagellonian University\\
Reymonta 4, PL 30-059 Cracow, Poland}\\
\mbox{ }\\
\end{center}

\vfill  

\begin{abstract}
It is shown that the gluons produced  
with small transverse momenta $k_\perp$ %
inside a jet ($k_\perp<k_\perp^{cut}$, $k_\perp^{cut} \to 0$) 
are independently emitted 
 from the primary parton, as QCD coherence suppresses their showering. 
Consequently, the low $k_{\perp}$ gluons follow a Poisson distribution,
very much like the soft photons radiated by a charged particle in QED.
On the contrary, the distribution of 
gluons with limited absolute momenta  $|\vec{k} | <k^{cut}$ 
remains non-Poissonian even for small $k^{cut}$.
It will be interesting to find out to what extent this perturbative prediction
for partons survives the  hadronization process. 
\end{abstract}

\newpage

\section{Introduction}



  It is by now well established that the multiplicity distributions of hadrons
produced in high energy collisions are substantially wider than the Poisson
distribution. This can be understood in a large class of models based on
branching processes \cite{polyakov}, the quark gluon cascade
in perturbative QCD being a relevant example \cite{kuv,bcm}.
In this paper we point out that the specific
properties of the QCD cascade allow one to select simple regions of phase space 
where the produced partons are less correlated and show a narrow 
multiplicity distribution. 
The experimental check of our predictions would
therefore provide a novel test of the role of perturbative QCD in multiparticle 
production.
   
It is convenient to characterize the multiplicity distribution 
of particles in an event by its global factorial moments 
$f^{(q)}=  
<n(n-1)\ldots (n-q+1)>$ of order $q$, or, written 
in terms of the $q$-particle inclusive
density
\begin{equation}
f^{(q)}(Q)=\int \rho^{(q)}(k_1\dots k_q;Q) d k_1\dots dk_q,
\label{global}
\end{equation}
with an integral over
the full phase space available at  the energy scale $Q$ for the process. 
 The ratios $F^{(q)}\equiv f^{(q)}/\bar{n}^q$ 
 were extensively studied in perturbative QCD
to various
degrees of accuracy \cite{dfk,mw,imd}. 
The asymptotic
results are obtained in the Double Logarithmic Approximation (DLA) and 
the normalized moments 
approach a limit $F^{(q)}(Q)\rightarrow F_q$, corresponding to KNO
multiplicity scaling\cite{polyakov,kno}. The numbers $F_q$ are  
independent of the coupling, thus providing a universal
characteristic of the multiplicity distribution. 
These moments essentially differ from $F_q=1$ as for
the Poisson distribution and then the asymptotic KNO distribution 
essentially differs from the Dirac delta function.

  In analogy with (\ref{global}) we define the {\em cut} moments
\begin{equation}
f^{(q)}_c(C,Q)=\int_{\Gamma_c(C,Q)} \rho^{(q)}(k_1\dots k_q;Q)
d k_1\dots d k_q , \label{cut}
\end{equation}
where the phase space integration is now restricted by $\Gamma_c(C,Q)$
with a cut variable $C$. In this 
paper we discuss two types of cuts to be applied to all particles in the
final state: 
the momentum cut $|\vec{k}_i| < k^{cut}$,
and the transverse momentum cut $k_{\perp,i} < k_{\perp}^{cut}$, 
for $i=1,...,q$.
Clearly, the moments (\ref{cut}) determine the multiplicity distribution
of particles produced in the restricted phase space, and as such  they
provide a more differential characteristics than the global quantities. 
For the maximal $k^{cut}$ and $k_\perp^{cut}$ at a given energy scale $Q$ the 
global quantities are retained.
    
    Depending on the choice of the cut, the moments in Eq.(\ref{cut})
 are probing quite
different physical phenomena. Of special importance in this discussion 
is the coherence of the soft gluon emission from the harder partons
which yields an angular ordering 
of subsequent emissions along the branching process \cite{efm}. 
In particular, this property leads to the suppression 
of the soft particle production \cite{adkt} and to a flat, energy
independent rapidity distribution of particles with low
transverse momenta in the parton jet, 
reminiscent of the QED Bremsstrahlung  \cite{klo}. 

To see the effect of the angular ordering on  
our cut moments consider first the momentum cut.
Since the dominant source of particle production in the QCD 
cascade is the gluon splitting,
the existence of one gluon simply enhances
the probability to find another one nearby in momentum space 
and this yields positive correlations
which do not vanish even for a small momentum cut $k^{cut}$. Such correlations would 
 also persist with the $k_{\perp}$ cutoff if there was no angular ordering.
 With the angular ordering
taken into account however, a cut in the transverse momentum of
one parton yields an even stronger restriction of
the available phase space for the next parton in the cascade,
 and consequently, the correlations are strongly damped near the 
low $k_{\perp}$ threshold. 
Because of this suppression of secondary emissions
in the $k_{\perp}$ cut phase space one obtains ultimately the independent
emission of low $k_{\perp}$ gluons and therefore a Poisson distribution 
very much in analogy to the
multiple soft photon Bremsstrahlung in QED.

     Differences between the momentum cut and $k_{\perp}$
cut moments become  more pronounced towards small cutoffs. 
Fortunately, this is the
region where the simple approximations are most reliable.
We derive here the evolution equations for both types of moments, 
 which  can be solved recursively to any
 order of the perturbation theory
within the DLA. These results provide
 the qualitatively correct picture as can also be inferred from 
 the Monte Carlo simulations at the parton level.

It will be interesting to find out whether the experimental data indeed show
the approach to a Poisson distribution as 
one might expect from the Local Parton Hadron Duality (LPHD) picture,
originally proposed for single inclusive distributions \cite{adkt}, but
successful also in a broader range of phenomena \cite{ko}.
Alternatively, the hadronization process could destroy the
coherence properties of the parton cascade; this happens  
in a specific hadronization model studied here for small
$k^{cut},k^{cut}_\perp<1$ GeV.
Hence the observation of the above regularities in the hadronic final state
would provide us with a clue on the hadronization mechanism.

After presenting a few basic relations of the DLA scheme in Chapter 2, we
derive in Chapter 3 the threshold behaviour of both families of 
cut moments in the leading perturbative order. In Chapter 4 recursive
evolution equations are derived which generate perturbative expansions
to arbitrary order for constant and running $\alpha_s$. In Chapter 5
we discuss for comparison the Monte Carlo results and conclude with Chapter 6.
The Appendices contain details of various derivations and a sample
 of explicit expressions for cut moments up to the third non-leading 
correction.

\section{Particle correlations in jets in DLA}


In this paper we consider multiplicity distributions  of hadrons or partons
in a jet of primary energy $P $ and half opening angle $\Theta $ 
in limited regions of phase space. The multiparton correlations in a jet
evolving from the primary parton~$a$ ($a=q,g$) can be derived from the
generating functional $Z_{P,a}$ which in DLA 
obeys the following integral
equation \cite{dfk}
\begin{equation}
   Z_{P,a}\{u\}=
\exp \left( \int_{\Gamma_P(K)} {\cal M}_{P,a}(K) [u(K) Z_{K,g}\{u\} -1]
d^3 K \right).
\labl{mez}
\end{equation}
This equation refers to the multiparton final state excluding the leading
particle. In (\ref{mez})
 the subscript $P$ denotes collectively the momentum vector of the
parent parton and
the half opening angle ($P=\{\vec P,\Theta\}$) of the jet it generates;
$\Gamma_P(K)$ stands for the phase space of the intermediate parent
$\vec{K}$,
  $\Gamma_P(K)=\{K : K<P, \Theta_{KP}<\Theta,K\Theta_{KP}>Q_0\}$,
where $Q_0$ is a transverse momentum cutoff
parameter). The probability ${\cal M}_{P,a}(K)$
for the bremsstrahlung of a single gluon
off the primary parton $a$ 
 reads for small angle $\Theta_{PK}$
\begin{equation}
 {\cal M}_{P,a}(K)d^3 K=c_a \gamma_0^2(K\Theta_{PK})\frac{dK}{K}
\frac{d\Theta_{PK}}{\Theta_{PK}}\frac{d\Phi_{PK}}{2\pi},
\labl{adef}
\end{equation}
where $c_g=1,~c_q=C_F/N_c=4/9$ for initial gluons and quarks respectively.
The multiplicity anomalous dimension is related to
the strong coupling by 
$\gamma_0^2 = 2N_C\alpha_s/\pi $ and
is taken as 
$\gamma_0^2 (y) = \beta^2 /(y+\lambda)$ where
$y = \ln (k_\perp/\Lambda)$, 
$\lambda= \ln \frac{Q_0}{\Lambda}$
and $\beta^2 = 4N_C/(\frac{11}{3}N_C - \frac{2}{3}n_f)$;
here $N_C$ and $n_f$ denote the numbers of colours and flavours and 
$\Lambda $ is the QCD scale.

The inclusive densities are obtained from Taylor expansion 
of the generating functional around $u(k)=1$
\begin{equation}
\rho_{P,a}^{(q)} (k_1,...,k_q)= \delta^q Z_{P,a}\{u\}/\delta u(k_1)...
\delta u(k_q)\mid_{u=1},      \labl{den}
\end{equation}
likewise the
connected (cumulant) correlation functions from
\begin{equation}
\Gamma_{P,a}^{(q)} (k_1,\ldots,k_q)= \delta^q \ln Z_{P,a}\{u\}/\delta u(k_1)\ldots
\delta u(k_q)\mid_{u=1} .     \labl{con}
\end{equation}
%
These differential distributions  
satisfy the following integral equations
\cite{ow}
 \begin{align}
\Gamma^{(q)}_{P,a}(k_1, \ldots ,k_q)& = 
d_{P,a,nest}^{(q)}(k_1, \ldots ,k_q) +
\int d^3K {\cal M}_{P,a}(K) \rho^{(n)}_{K,g} (k_1, \ldots ,k_q), 
\labl{gamman}\\
d_{P,a,nest}^{(q)} (k_1, \ldots ,k_q)& = 
{\cal M}_{P,a}(k_1)\rho_{k_1,g}^{(q-1)}
(k_2, \ldots, k_q) + cycl.               
\labl{dnest}
 \end{align}
The cut moments  $f^{(q)}_c$ are obtained 
as integrals over the particle density
distributions $\rho^{(q)}$ as in eq.~\eref{cut}, likewise 
the cumulant moments $c^{(q)}$ as integrals over the
connected correlation functions $\Gamma^{(q)}$. 

The global moments integrated over full phase space can be 
derived more simply from the integrated generating function $Z(Y,u)$
which -- in analogy to eq.~\eref{mez} --
fulfils the evolution equation \cite{dktm} 
\begin{equation}
Z_a(Y,u) = \exp \{ \int^Y_0 dy c_a \gamma^2_0(y)(Y-y)[uZ(y,u)-1]\}
     \labl{Zdla}
\end{equation} 
where the evolution variable  in small angle approximation is 
$Y = \ln \frac{P\Theta}{Q_0}$.
The global moments are then derived 
from $f^{(q)} = d^qZ/du^q|_{u=1}$ and $c^{(q)} = d^q\ln Z/du^q|_{u=1}$
and we recall that these moments  are related by (see, for example \cite{wdk})
\begin{equation}
f^{(q+1)} = \sum^q_{m=0} {q \choose m} f^{(q-m)} c^{(m+1)}.      \labl{fcrec}
\end{equation}
Frequently used are the moments $F^{(q)} = f^{(q)}/\bar{n}^q$ and 
$K^{(q)} = c^{(q)}/\bar{n}^q$ normalized by the mean multiplicity
$\bar{n} \equiv f^{(1)} \equiv c^{(1)}$. 

For later use we also present here the results for
the global factorial moments $f^{(q-1)}$  to the 
lowest perturbative order, an approximation suitable for small $Y$.
To this end, one obtains first, 
after q-fold differentiation of $\ln Z$ over $u$ in (\ref{Zdla}),
the equation
\begin{equation}
c_a^{(q)}(Y) = \int^Y_0 dy c_a \gamma^2_0(y)(Y-y)(qf^{(q-1)}(y)+f^{(q)}(y)).
       \labl{fcmom}
\end{equation}

Then, starting with 
$f^{(0)}$ = 1, one finds the higher moments by iteration
 whereby in leading order of $Y$ the second term with $f^{(q)}$ 
in (\ref{fcmom}) can be neglected
against $qf^{(q-1)}$. After 
inserting (\ref{fcmom}) into (\ref{fcrec}) one obtains finally for 
small $Y$
\begin{equation}
f^{(q)}(Y) \simeq f_q c_a^q \gamma_0^{2q}Y^{2q} ,          \labl{fqborn}
\end{equation}
thus the moments vanish at the threshold $Y\to 0$ ($P\Theta \to Q_0$).
The coefficients $f_q$ in (\ref{fqborn})
 are given recursively by 
 \begin{equation}
f_q = \sum^{q-1}_{k=0} \frac{f_kf_{q-k-1}}{2(2q-2k-1)}{q-1 \choose k}
     \labl{fqrec}
\end{equation}
\noindent
and the first coefficients read
\begin{equation}
f_0 = 1,\quad f_1 = \frac{1}{2},\quad f_2 = \frac{1}{3},\quad
 f_3 = \frac{17}{60}, \ldots .
     \labl{fqfirst}
 \end{equation}
Note that the 
 relation (\ref{fqrec}) is different from the analogous
relation which determines global moments at large $Y$ (\cite{dktm,ow}).

\section{Multiplicity Moments for small cut-off} 

In this section we derive the multiplicity moments for the 
phase space with cylindrical cut
$k_\perp< k_\perp^{cut}$ and with  spherical cut 
$k<k^{cut}$ for small parameters $k_\perp^{cut}$ and $k^{cut}$. 
In these limiting cases various simplifications can be
applied: first, it is enough to take into account the terms of 
lowest order in the coupling. We show in the next section 
 that terms of higher order are 
indeed suppressed for small transverse momenta. Secondly, in this limit,  
the solutions for  running coupling approach those for fixed
coupling, in which case 
close expressions can be
obtained.

\subsection{Cylindrical phase space}

For the multiplicity in the $k_\perp^{cut}$ cylinder 
(not including the leading particle of momentum $P$) we obtain
\begin{equation}
\bar n (k_\perp^{cut}, P, \Theta) = \int^{k_\perp^{cut}}_{Q_0}
\frac{dk_\perp}{k_\perp} \gamma^2_0 (k_\perp) 
\int^{\Theta}_{k_\perp/P} \frac{d\Theta'}{\Theta'} = 
\int^{X_\perp}_0 dy\gamma_0^2(y)(Y-y) ,      \labl{mult}
 \end{equation}
\noindent
where $X_\perp = \ln \frac{k_\perp^{cut}}{Q_0}$. Now it can easily be seen that for
 $X_\perp \ll \lambda$  the variation of the coupling can be neglected and
 $\gamma_0^2=\beta^2/\lambda$.
For fixed coupling we find in lowest order 
\begin{equation}
\bar n(X_\perp,Y) = \gamma^2_0 X_\perp (Y-X_\perp /2),        \labl{multfix}
\end{equation}
which reduces for $X_\perp\ll Y$ to $\bar n(X_\perp,Y) = \gamma^2_0 X_\perp 
Y$ corresponding to
a constant particle density in $\ln k_\perp$ and $\ln k$. 

Here and in the rest of this section 
we write for simplicity the results for
gluon jets. For quark jets the coupling $\gamma_0^2$ should be replaced by 
$c_q \gamma_0^2$ everywhere as we deal only with gluon emission from the
primary parton.

The higher order moments are obtained from (\ref{gamman}), where we neglect
the second term which is of higher order in $\alpha_s$.
The integration of $\Gamma^{(q)}$ and $\rho^{(q)}$ over the momenta yields 
the cumulant and factorial moments in the given region of phase space. 
For $k_1$ kept fixed the integral over $k_2 \cdots k_q$
in (\ref{dnest}) yields the global factorial moment $f^{(q)}$
at scale $k_{\perp 1} = k_1\vartheta_1$:
because of angular ordering the angles in all later emissions must obey
$\vartheta_2, \cdots, \vartheta_q < \vartheta_1$.
So one finds
 \begin{equation}
c^{(q)}(X_{\perp},Y) = q \int^{X_{\perp}}_0 dy \gamma^2_0(y)(Y-y)f^{(q-1)}(y).  \labl{cqeq}
 \end{equation}
Hence in this approximation the cut moments are determined by the
 global ones.
Again one recovers the fixed $\alpha_s$ limit for small $X_\perp$.


For the normalized cumulant moments in the case of fixed coupling we find from
 (\ref{multfix}) and (\ref{cqeq}) after inserting the moments in the
approximation (\ref{fqborn})
\begin{equation}
K^{(q)}(X_\perp,Y) = \frac{2^{q-1}f_{q-1}}{2q-1} 
          \frac{2qY-(2q-1)X_\perp}{(2Y-X_\perp)^q} X_\perp^{q-1},
        \labl{KqLO}
\end{equation}
\noindent
and for the case of small cut-off $k^{cut}_\perp$ we are interested in,
 i.e. for $X_\perp \ll Y$
\begin{equation}
K^{(q)}(X_\perp,Y) = 
\frac{qf_{q-1}}{2q-1} \left(\frac{X_\perp}{Y}\right)^{q-1}. \labl{Kqsx}
\end{equation}
Therefore, in this approximation the cumulants $K^{(q)}$ quickly
decrease with the order $q$. 
It is interesting to note that this power like 
dependence on the order $q$ corresponds to the
\lq\lq linked pair" ansatz for correlations \cite{ces}, usually written as
$K^{(q)} = A_q [K^{(2)}]^{q-1}$, which is made  
in models where multi-particle correlations
are built from 2-particle correlations. The higher cumulants 
in (\ref{Kqsx}) rise more
slowly than in case of the negative binomial distribution in which case
$A_q=(q-1)!$ \cite{DeW}.

Finally, the factorial moments are 
obtained by solving (\ref{fcrec}) approximately through
$F^{(q)} = 1 + 
q(q-1)K^{(2)}/2 + {\cal O} (K^{(3)})$, 
which  yields
\begin{equation}
F^{(q)}(X_\perp,Y) \cong 1 + \frac{q(q-1)}{6} \frac{X_\perp}{Y}.
       \labl{FqLO}
\end{equation}

So we obtain the remarkable result that for small transverse momentum
cut-off all factorial moments approach unity and therefore the multiplicity
distribution becomes Poissonian. This is a consequence of the dominance of
the single soft gluon emission at small $k_\perp$, i.e., the absence of
branching processes with secondary gluon emissions. This behaviour is
just analogous to the usual QED bremsstrahlung and follows from the
coherence of the soft gluon radiation and the angular ordering
condition which limits the angles of the secondary particles in
(\ref{cqeq}) by the (typically small) emission angle $\vartheta_1$
of the first gluon.

If the running of the coupling is taken into account, 
the computation already of the leading contributions yields quickly rather
lengthy formulae with increasing order $q$. For the first two moments
we find 
\begin{gather}
\bar n(X_\perp,Y) = \beta^2\{(Y+\lambda)\ln \frac{X_\perp  + \lambda}{\lambda}
-X_\perp \},
           \labl{multrun}\\
\begin{align}
 c^{(2)}(X_\perp,Y)&=\beta^4\{[\lambda(4Y+3\lambda)-X_\perp (X_\perp -2Y)]
         \ln \frac{X_\perp +\lambda}{\lambda}\nonumber\\
    &\phantom{=\beta^4\{[\lambda(4Y+3\lambda)}
         -X_\perp (3\lambda+4Y-3X_\perp /2)\}\labl{c2runful}\\
 &\approx 2\beta^4Y\{(X_\perp +2\lambda)\ln \frac{X_\perp
+\lambda}{\lambda}-2X_\perp \},
            \labl{c2run}
\end{align}
\end{gather}
where the latter approximation applies for $X_\perp  \ll Y$.
Our essential results about the behaviour 
of correlations for $X_\perp  \to 0$
coincide with the results for fixed coupling 
as expected by the general argument. 

The different approximations for the second moment $F^{(2)}$ discussed here are
compared in Fig.~\ref{figtheory1}. One can see that the running $\alpha_s$ 
results yield lower values for the moments with increasing $X_\perp$ but all
results approach the same linear behaviour with the same slope for 
$X_\perp \to$ 0.

It is also interesting to note that the results (\ref{KqLO}-\ref{FqLO})
for the normalized moments are independent of the coupling $\gamma_0^2$ and
therefore also independent of the colour of the primary parton, i.e. the
normalized moments in quark and gluon jets approach the same limiting
behaviour. The reason is that in this limit all gluons are emitted from the
primary parton; this is different from the case of global moments, where
only the first gluon in the cascade is emitted from the primary parton within
the DLA.

\subsection{Spherically cut phase space}

Next we consider the multiplicity distribution in the reduced phase space
with cut in the modulus of momentum  $k<k^{cut}$, but still 
limited by the jet opening angle $\Theta$. 

In the present approximations the initial momentum $P$ enters the final
results only through the boundary $k<P$ if the leading particle is
disregarded. 
This implies that we can easily construct from the previous subsection the
results for the combined cuts  $k_\perp<k_\perp^{cut}$ and $k<k^{cut}$
by replacing $P$ by $k^{cut}$ or 
  $Y$ by $X=\ln (k^{cut}\Theta/Q_0)$ in the equations for moments such 
as~\eref{FqLO}. We note that 
the largest deviations from the DLA are expected to come from the
large momenta $k$ where the 
energy-momentum constraints and also the explicit form of the splitting
functions become important. Therefore, 
a DLA formula like (\ref{FqLO}) becomes more 
realistic for restricted momenta $k < k^{cut}$, i.e., if we substitute 
$X$ for $Y$ keeping $X$ smaller than the global scale $Y$ at the same time. 

Now the results for the spherical phase space with $k<k^{cut}$ 
(without additional bound on $k_\perp$) are obtained 
from the formulae of the previous subsection 
after letting $k_\perp^{cut}\to k^{cut}$
or  $X_\perp \to X$ and also replacing $Y$ by $X$. 
For fixed coupling in lowest order we obtain from (\ref{KqLO})
\begin{align}
\bar n(X,Y) &= \gamma_0^2 X^2/2 ,   \labl{nbark}\\
K^{(q)}(X,Y)& = 2^{q-1} f_{q-1}/(2q-1). 
\labl{Kk}      
\end{align}      
The normalized moments $K^{(q)}$ are 
$X$-independent. This shows the different
behaviour of the cut moments in cylindrical and spherical 
phase space: 
they decrease to zero or stay constant, respectively, 
if the cut-off is lowered
towards the kinematic limit.
Consequently, as announced in the Introduction, soft gluons with limited
momentum $k$ 
have essentially a non-Poissonian multiplicity distribution, while
those with limited $k_{\perp}$ are indeed produced independently.

\subsection{Role of angular ordering}

In order to demonstrate the 
importance of angular ordering for these predictions,
 we arbitrarily drop 
this condition in the corresponding integrals. Let's
consider first the integral as in 
(\ref{cqeq}) in the case of the cylindrical phase
space. 
Taking the second moment as an example,
we then  obtain  in the approximation $X_\perp \ll Y$ and $\alpha_s$ fixed
\begin{align}
c^{(2)}(X_\perp,Y)& = 2 \int^{\Theta}_{Q_0/P} \frac{d\vartheta_1}{\vartheta_1}
 \int^{k_\perp^{cut}/\vartheta_1}_{Q_0/\vartheta_1} 
\frac{dk_1}{k_1} \gamma^2_0
 \int^{\Theta}_{Q_0/k_1} \frac{d\vartheta_2}{\vartheta_2}
 \int^{k_\perp^{cut}/\vartheta_2}_{Q_0/\vartheta_2} \frac{dk_2}{k_2}
\gamma^2_0\nonumber\\
 &\approx \gamma_0^4 X_\perp^2 Y^2
      \labl{c2no}
\end{align}
\noindent
where the upper limit of the $\vartheta_2$-integral is shifted from 
$\vartheta_1$ to $\Theta$.
The result is symmetric in $X_\perp$
and $Y$ as is the corresponding result for the multiplicity (\ref{multfix})
which applies also to the present discussion; then the normalized moment 
becomes constant
\begin{equation}
K^{(2)} \approx 1.  \labl{K2no}
\end{equation}
On the other hand, if angular ordering is taken into account, 
one obtains the asymmetric result
(\ref{Kqsx}) -- corresponding to $c^{(2)} \sim X_\perp^3 Y$ with the
stronger $X_\perp$ dependence --
and finally the Poisson distribution in the limit $X_\perp \to 0$.

Of course, angular ordering has also consequences for the results in 
the spherical phase space and in general reduces the correlations,
for example, $K^{(2)}=\frac{1}{3}$ from (\ref{Kk}) as compared 
to  (\ref{K2no}), however its
consequences are less dramatic in this case. 
For the $k_\perp$ cut the primary
emission angle is restricted towards small angles 
and then the same holds for the subsequent emission angles;
in the case of spherical cut the primary emission angle is only
limited by its maximum, the  jet opening angle $\Theta$, and therefore
the restrictions for the subsequent emissions are less severe.
So the $k_\perp$ cut results are more sensitive to the constraints from
coherence and angular ordering and therefore yield the very different results.

\subsection{Alternative phase space boundaries}

Finally in this section, we want to discuss a different upper bound for
the momenta which connects to other results. In case of cylindrical cut 
we have considered so far the trapezoidal
boundaries $Q_0 \leq k_\perp \leq k_\perp^{cut}$ and
$k_\perp/\Theta < k < P$. In this case $X_\perp \leq Y$ where 
$X_\perp = Y$ corresponds to the triangular boundary
of the spherical cut. For comparison we consider now 
a phase space boundary which is given by a second cone with opening angle 
$\Theta' < \Theta$. In this case  the upper limit 
of $k$, at fixed $k_\perp$, becomes  $k < k_\perp/\Theta'$. 
Then the angular integrals factorize and, instead of (\ref{multfix})
and (\ref{KqLO}) one finds simply
\begin{align}
\bar n(X_\perp, \Delta Y) &= \gamma_0^2 X_\perp \Delta Y    \labl{multang}\\
K^{(q)}& = \frac{q f_{q-1}}{2q-1} \left(\frac{X_\perp}{\Delta Y} \right)^{q-1}
\labl{Kang}
\end{align}
where $\Delta Y = \ln(\Theta/\Theta')$ denotes the rapidity difference
of the angular boundaries. These results again clearly demonstrate the
different roles of longitudinal and transverse directions: for
$X_\perp  \ll \Delta Y$ the correlations vanish because of angular ordering
as discussed above and the
formulae (\ref{multang}) and (\ref{Kang}) correspond to  the results 
(\ref{multfix}) and (\ref{Kqsx}) for small $X_\perp$ with $Y$ replaced by $\Delta
Y$; on the other hand,  
for $X_\perp  \gg \Delta Y$ the collection of the many produced partons
yields large correlations. In fact, in this configuration the higher orders 
become important with increasing $X_\perp$ 
and the fractal structure of the parton cascade yields
 the power behaviour of the
correlations (``Intermittency'') \cite{ow,dd,bmp}. 
Eq. (\ref{Kang}) nicely demonstrates the different trends $K^{(q)}\to 0$ and
$K^{(q)}\to \infty$ for $X_\perp\to 0$ and $\Delta Y\to 0$ respectively.
\section{Multiplicity moments in higher orders}
  In  this section we derive 
 the integral evolution equations 
for factorial moments in the cut phase space. 
These equations will then be used to generate in the 
systematic way higher order corrections to the required accuracy. 
The complete solution obtained in this way gives cut moments for the
full range of the cut-offs $0 <X,\xt< Y$. It reduces to the Born approximation
at the threshold, thus justifying simplifications used in the previous Section.
For simplicity, only the gluon jets will be considered in this section,
hence we set $c_a=1$. 

 All evolution  
equations, which we seek for, follow from Eqs.(\ref{gamman},\ref{dnest}) 
together with~\eref{fcrec} which is also valid in
the restricted phase space. To begin we first simplify Eqs.(\ref{gamman})
employing the pole approximation \cite{ow}.
This essential step consists of
saturating the angular integration in (\ref{gamman}) 
over the intermediate parent momentum
$\vec{K}$ by the leading singularities of the integrand. This amounts to
approximating $\vec{K}\parallel\vec{k_i}$ in all nonsingular terms and
integrating only over the smallest angle $\Theta_{Kk_{i}}$, $k_i=1\dots n$.
This leads to the following equation for the single parton 
density differential in momentum $k$ and angle $\vartheta$ \cite{ow}
\begin{equation}
\rho^{(1)}(\vartheta,k,P)  = {\ad(k\vartheta)\over k\vartheta}
  + \frac{1}{\vartheta}
  \int_{k}^P {dK\over K} \int_{Q_0\over k}^{\vartheta}
  {d\Theta_{Kk}\over\Theta_{Kk}} \gamma_0^2(K\vartheta)
  \left[\Theta_{Kk}\rho^{(1)}(\Theta_{Kk},k,K)
  \right],
  \labl{tkk}
\end{equation}
For the fully differential two-parton connected correlation function
\begin{equation}
\Gamma_P^{(2)}(\vartheta_1,\vartheta_2,k_1,k_2)=
\Gamma_1^{(2)}(\vartheta_1,\vartheta_{12},k_1,k_2,P)
+\Gamma_2^{(2)}(\vartheta_2,\vartheta_{12},k_2,k_1,P), \label{spli}
\end{equation}
 we 
obtain
\begin{eqnarray}
\lefteqn{
\Gamma^{(2)}_1(\vartheta_1,\vartheta_{12},k_1,k_2,P)=
{\cal M}_P(k_1)\rho^{(1)}(k_2,\vartheta_2,K) + 
\frac{1}{\vartheta_{1} }
\int_{k_{>}}^P \frac{dK}{K} \gamma_0^2(K\vartheta_1)} \nonumber \\ &&
\left\{\int_{Q_0/k_1}^{\vartheta_{12}}d\Theta_{Kk_1}
\rho^{(1)}(\Theta_{Kk_1},k_1,K)\rho^{(1)}(\vartheta_{12},k_2,K)+
\right.  \nonumber \\ & &
\left.\int_{\vartheta_{12}}^{\vartheta_{1}} d\Theta_{Kk_1}
\Gamma^{(2)}_1(\Theta_{Kk_1},\vartheta_{12},k_1,k_2,K) \right\}. \labl{g2diff}
\end{eqnarray}
with the polar angle $\vartheta_i$ of particle $i$,  the relative polar angle
$\vartheta_{12}$ of two particles and $k_> = \max \{k_1,k_2\}$. 
Note the different bounds on $\Theta_{Kk_1}$ in the 
two terms. They follow from the
different singularity structure of the product term and of the connected
correlation function. The splitting (\ref{spli}) is a consequence of the 
pole approximation -- $\Gamma^{(2)}_i$ results from the saturation of the 
$d^3K$ integration in Eq.(\ref{gamman}) by the $\vec{K}\parallel\vec{k}_i$
configuration.

Analogous simplifications of the evolution equations can be made for higher
order densities. 

\subsection{Evolution equations for momentum cut moments}

Integrating Eq.(\ref{tkk}) over the spherically cut phase space of the
single parton in a jet $(P,\Theta)$ gives for the average multiplicity
\begin{equation}
\bar{n}(k^{cut},P\Theta)  = \bar{n}(k^{cut}\Theta)+
\int_{Q_0/k^{cut}}^{\Theta} {d\vartheta\over\vartheta}  
\int_{k^{cut}}^{P} {dK\over K} \ad(K\vartheta) \bar{n}(k^{cut},K\vartheta),
  \labl{nmom}
\end{equation}
The structure of this equation can be simply understood from the general 
rules of the
angular ordering and transverse momentum limitations. Since  
the momentum of a child parton is limited by $k^{cut}$, 
 their emission angles
are bounded from below by $Q_0/k^{cut}$. Because of the angular ordering, 
this also limits from below the emission angle $\vartheta$ of the intermediate
parent $K$.
The remaining phase space of $(K,\vartheta)$ can be conveniently split into
$K<k^{cut}$ and $K>k^{cut}$, c.f. regions (I) and (II) in Fig.~\ref{fkin1}. 
Parents in the first region contribute to the
global multiplicity at the scale $k^{cut}$ and parents from the second
region give rise to the second term of Eq.~(\ref{nmom}). In Appendix A we
derive an analogous equation for the second cumulant, by following the
 above steps in detail.

Similarly, one derives the evolution equations for higher moments. We obtain
for the cumulants $c^{(q)}(\kc,P\Theta)=c^{(q)}(X,Y)$ in the logarithmic
variables $X=\ln{(\kc\Theta/Q_0)}$ and $Y=\ln{(P\Theta/Q_0)}$ 
\begin{eqnarray}
\lefteqn{c^{(q)}(X,Y) =  
\int_0^X dy \ad(y)[qf^{(q-1)}(y)+f^{(q)}(y)](X-y) }
\nonumber \\&&+
\int_0^X dx\int_x^{Y-X+x} dy \ad(y)f^{(q)}(x,y) .
\labl{cqmom}
\end{eqnarray}
This, together with Eqs.(\ref{fcrec},\ref{fcmom}) , uniquely determines all multiplicity
moments in the spherically cut phase space. The lowest order contributions
are given recursively, in $q$, by the first integral with the $f^{(q-1)}$ 
term only. Perturbative expansion in the order $O(\ad^{k})$ can now be 
generated recursively in $q$ and $k$ by integrating the
 $O(\ad^{k-1})$ expansion according to Eq.(\ref{cqmom}). 
For constant $\alpha_s$ this was done for the
first four moments up to terms $O(\ad^{8})$ with the results shown in
Fig.~\ref{figtheory}.
 A sample of explicit expressions is given in the 
Appendix B.

   In the global limit $(X\rightarrow Y)$ the second integral disappears,
and the above equations reduce to the known equations (\ref{fcmom}) for
the global moments.

Since the lowest order is also given by the first integral, the Born
approximation for the normalized moments is independent of the maximum
virtuality $P\Theta$ and equals to the global moments at the scale
$\kc\Theta$. Consequently, the normalized moments are constant in this
approximation and are determined by the coefficients (\ref{fqrec}). 
In the higher orders they aquire a mild $X$ dependence
due to the additional radiation from parents which are faster than the 
cut-off $\kc$.

As already mentioned, while the higher order corrections are of course
important for the unnormalized moments, they largely cancel in the 
ratios $F^{(q)}$. As a consequence, the Born approximation discussed
in the previous chapter reproduces rather reliably the properties of these
moments, especially the difference between the spherically and cylindrically
cut moments. 
  
    This scheme applies  for the running $\alpha_s$ 
case as well, and higher order
corrections can be generated by the same, albeit more tedious steps. 
 
\subsection{Cylindrically cut moments}

\subsubsection{Multiplicity for constant $\alpha_s$}
    Integrating Eq.(\ref{tkk}) over the parton momentum $k$ at fixed $\kt$
gives the evolution equation for the inclusive $\kt$ distribution
\begin{equation}
\rho(\kt,P\Theta)=b(\kt,P\Theta)+{\ad\over\kt}
\int_{\kt/\Theta}^P {dK\over K} \int_{Q_0}^{\kt} d\kappa_{\perp}
\rho(\kappa_{\perp},K\Theta{\kappa_{\perp}\over\kt}) \labl{rhokt}
\end{equation}
with the Born term $b=\ad\ln{(P\Theta/\kt)}/\kt$. Note that the
kinematics of this equation is very different from that for the absolute
momentum distribution. 
While in the latter the rotationally invariant momentum of a child parton 
is the same for the P-jet and for the internal K-jet, in the former the
transverse momenta $\kt$ and $\kappa_{\perp}$ refer to {\em different}
axes, see Fig.~\ref{fkin2}. This leads to the more complicated 
structure of Eq.(\ref{rhokt}).
In particular, the virtuality of the internal K-jet depends now on
$\kappa_{\perp}$. In fact the effective angle 
$\bar{\vartheta}=\Theta \frac{\kappa_{\perp}}{\kt}$ should be interpreted as the
angle between the momentum of the parent $\vec{K}$ and the parton emitted at
the {\em  maximal} angle $\Theta$ which is allowed. In this variable,
Eq.(\ref{rhokt}) has a simple interpretation. 

   Integrating Eq.(\ref{rhokt}) over $\kt$ up to the transverse cut-off
$\kct$, we obtain for the average multiplicity of partons with limited $\kt$
\begin{eqnarray}
\lefteqn{\bar{n}(\kct,P\Theta)=\bar{n}_b(\kct,P\Theta)+} \nonumber\\&&
\ad\int_{Q_0/\Theta}^{\kct/\Theta} {dK\over K} 
\int_{Q_0/K}^{\Theta} {d\bar{\vartheta}\over\bar{\vartheta}} 
\bar{n}(K\bar{\vartheta})+ \nonumber\\&&
\ad\int_{\kct/\Theta}^{P} {dK\over K} 
\int_{Q_0\Theta/\kct}^{\Theta} {d\bar{\vartheta}\over\bar{\vartheta} } 
\bar{n}(\kct {\bar{\vartheta}\over\Theta},K\bar{\vartheta}). \labl{navkt}
\end{eqnarray}
where $\bar{n}_b$ denotes the Born contribution, Eq.(\ref{multfix}). Similarly to
the spherical case, the phase space of the intermediate parent is divided
into two regions. In the first integral the maximum virtuality of the
K-jet is limited by $\kct$, therefore all partons in that jet are below
cut-off, hence the global multiplicity contributes. In the second integral
some partons may exceed the external cut-off $\kct$, thus only cut
multiplicity contributes. The cut 
$\kappa_{\perp}^{max}=(\kct/\Theta)\bar{\vartheta}$ corresponds to the 
 hardest parton compatible with the cut-off $\kct$ emitted from the parent K
at angle $\bar{\vartheta}$.
  The above equation reproduces the exact solution which can also be 
obtained, in the case of constant $\alpha_s$,
by direct integration of the known 
fully differential distribution. 

\subsubsection{Relation between spherically and cylindrically cut moments}

   We have seen that the natural variable appearing in the $\kt$ cut case 
is the angle between the hardest parton in an original $(P,\Theta)$ jet and 
the intermediate parent $K$. One can however reorganize the calculation such 
that the phase space of the intermediate parton K is parametrized by its
momentum  and the angle $\Theta_{PK}$ at which it was emitted from the 
original parent $P$. This formulation provides very useful relations 
between both families of cut moments, valid also for running $\alpha_s$.
It will be used to generate cylindrically cut moments to 
higher orders.
   
   We begin again with the first moment, i.e. the average multiplicity
written as, c.f. Fig.~\ref{kin3} 
\begin{equation}
\bar{n}(\kct,P\Theta)=\bar{n}(\kct)+\int_{\kct/P}^{\Theta} d\vartheta
\int_{Q_0/\vartheta}^{\kct/\vartheta} dk \rho^{(1)}(\vartheta,k,P). \labl{nsplit}
\end{equation}
Due to the independence of the fully differential distribution
on the opening angle $\Theta$, the integration over the small angles of the
emitted partons $Q_0/P < \vartheta < \kct/P $ results in the global
multiplicity at the scale $\kct$ (region A in Fig.~\ref{kin3}). The second term
represents contribution from the region B.  
 We use now the Eq.(\ref{tkk}) to express
the last term  by the contributions from the intermediate
parents. Rearranging the integrals gives after some algebra 
\begin{eqnarray}
\lefteqn{\bar{n}_{cyl}(\xt,Y)= \bar{n}(\xt)+(Y-\xt)\int_0^{\xt} 
\ad(y)[1+\bar{n}(y)]dy}
\nonumber \\&&
+\int_{\xt}^Y (Y-y) \ad(y) \bar{n}_{sph}(\xt,y) dy . \labl{navrel}
\end{eqnarray}
where the subscripts distinguish different cuts, and logarithmic variables
are used. Here and in the following the quantities without subscripts and one
argument refer to the global quantities. 
This equation allows to calculate $\kt$-cut multiplicity if the
momentum cut result is known. Using the constant $\alpha_s$ solution of 
Eq.(\ref{nmom}) we have reproduced with the aid of (\ref{navrel})
perturbative expansion of $\bar{n}_{cyl}$ to $O(\ad^{12})$.

   This approach generalizes for the moments of arbitrary order.
We begin with the $q=2$ case and display explicitly the relevant two-parton 
phase space for the global cumulant
\begin{equation}
c^{(2)}(P\Theta)=2\int_{Q_0/P}^{\Theta}d\vartheta_1
\int_{Q_0/P}^{\vartheta_{1}} d\vartheta_{12}
\int_{Q_0/\vartheta_{12}}^P dk_1
\int_{Q_0/\vartheta_{12}}^P dk_2 
\Gamma^{(2)}_1(\vartheta_{1},\vartheta_{12},k_1,k_2,P), \labl{phsp2}
\end{equation}
where $\Gamma^{(2)}_1$ defined as in Eq.(\ref{spli}).
The factor $2$ comes from the symmetric $(1\leftrightarrow 2)$ configuration.
Fully differential connected correlation functions in the leading logarithmic 
accuracy depend only on {\em one} angle relative to the original parent 
momentum, e.g. $\vartheta_1$. Direction of the second particle  enters only 
via the relative angle $\vartheta_{12}$. Moreover for 
$\vartheta_{12}>\vartheta_1$ the connected correlation function 
vanishes - yet another manifestation of the angular ordering \cite{ow}.
Finally, the fully differential connected correlation function 
is independent of the
global opening of the original jet as in the single parton case.
The last property allows us to split the cut cumulant into two parts as
previously
\begin{eqnarray}
\lefteqn{c_{cyl}^{(2)}(\kct,P\Theta)=
c^{(2)}(\kct)+2\int_{\kct/P}^{\Theta}d\vartheta_1
\int_{\vartheta_1 Q_0 /\kct}^{\vartheta_{1}} d\vartheta_{12} }\nonumber \\&&
\int_{Q_0/\vartheta_{12}}^{\kct/\vartheta_1} dk_1
\int_{Q_0/\vartheta_{12}}^{\kct/\vartheta_1} dk_2 
\Gamma^{(2)}_1(\vartheta_{1},\vartheta_{12},k_1,k_2,P). \labl{c2split}
\end{eqnarray}
Since the leading singularities are generated from the
region where in fact $\vartheta_{12} \ll \vartheta_1$, the measured transverse
momentum of the second parton $k_{2\perp}\sim k_2\vartheta_1$ and this
gives the upper limit of the $k_2$ integration in (\ref{c2split}).
Inserting now Eq.(\ref{g2diff}) under the integral in (\ref{c2split}) and 
 rearranging various integrals gives the final expression for the cylindrically cut 
cumulants in terms of the global and the spherically cut factorial moments.
Details of this calculations are presented in Appendix C.

Similar steps allow to derive analogous relations for cumulants of
arbitrary order $q$. We obtain  
\begin{eqnarray}
\lefteqn{ 
         c^{(q)}_{cyl}(\xt,Y)= c^{(q)}(\xt)+} \nonumber \\&&
(Y-\xt)\int_0^{\xt} \ad(y)
[qf^{(q-1)}(y)+f^{(q)}(y)]dy+ \nonumber \\&& 
\int_{\xt}^Y (Y-y) \ad(y) f^{(q)}_{sph}(\xt,y) dy . \labl{cqrel}
\end{eqnarray}
  In fact Eq.(\ref{navrel}) is the special case of this relation for $q=1$
since $f^{(0)}=1$. 

  Together with Eqs.(\ref{cqmom}) this equation uniquely determines 
all moments
with limited $k_{\perp}$. We have used these equations to generate
algebraically the perturbative expansion of the first four moments up to 
the eight order in $\ad$ for constant $\alpha_s$, see Fig.\ref{figtheory} and Table 4. 
We quote only the
first nontrivial example - the $O(\alpha_s^3)$ correction for $c^{(2)}$
\begin{equation}
 c^{(2)}_{cyl,3}(\xt,Y)={7\over 72}\xt^6-{1\over 4} \xt^5 Y + 
{1\over 6} \xt^4 Y^2 . \labl{c23}
\end{equation}
This result was confirmed independently by the direct integration of
the resolvent representation for $\rho^{(2)}$ \cite{ow} and also from the
first explicit iteration of Eq.(\ref{gamman}). 

\subsection{Properties of the higher order results}

 Fig.~\ref{figtheory} 
shows spherically and cylindrically cut moments up to $O(\alpha_s^8)$ 
 in the full range of the cut-off. First three normalized moments are displayed.
All properties discussed earlier are fully confirmed by this calcuation.
    For small cutoffs all cylindrically cut moments tend to unity while those with limited
momentum have different limits corresponding to the non-Poissonian multiplicity distribution
of soft gluons.  
        The perturbative expansion is rather poorly convergent for the unnormalized 
moments which, being polynomials in the cut-off and $Y$, grow faster for higher order 
$k$. 
However this growth is largely canceled in the ratios. 
As a consequence, perturbative expansion is remarkably stable for the normalized moments $F^{(q)}$.  
This  is clearly seen in Fig.~\ref{figtheory}
where different lines, corresponding to different
maximum order  $k_{max}$ included,   group
 together for each $q$ and each type of the cut-off.
In the global limit $\xt,X=Y$ spherically and cylindrically cut  moments are equal. 
However this common value, which is in fact given by the global 
moment, {\em does} vary with the order of the perturbative approximation. 
This is seen in Fig.~\ref{figtheory} as well. 
This dependence is very weak and corresponds, for large $Y$, 
to the difference between the coefficients which 
determine the threshold behaviour (see Eq.(\ref{fqrec}))   
and the asymptotic behaviour  
(see \cite{ow}), respectively, of the global moments.


\section{Monte Carlo results}


We have studied the behaviour of the 
cut moments within a Monte Carlo program in order to check our analytical
calculations with a more complete numerical method, which is important in the
test of the LPHD picture. 
 In view of the similarity of its scheme with our analytical framework,
we have chosen the ARIADNE program\cite{ariadne},
in which the perturbative phase is terminated by a cutoff in transverse
momentum and, in its default version, 
followed by string fragmentation. In deviation from the default value, 
the cutoff $Q_0$ for partons can be
chosen close to the QCD scale $\Lambda$ as required 
in the LPHD picture, which is not easily possible in 
 other popular QCD Monte Carlo models. 

For our applications we have then switched off the
hadronization phase within ARIADNE and  directly compared 
the result at the end of the perturbative evolution with our 
theoretical predictions. 
We have reset the values of the parameters $Q_0$ and 
$\Lambda$ in ARIADNE in order to directly reproduce the energy dependence of
the average multiplicity of all (charged plus neutral) hadrons according to
the LPHD picture. We  estimated 
the full (charged plus neutral) multiplicity 
as  3/2 the measured multiplicity for charged hadrons and 
 we fixed the overall
normalization factor which relates partons to all 
hadrons to $K_{all}$ = 1, as suggested by previous studies on the
average multiplicity\cite{lomult}\footnote{There is a correlation among 
the parameters $Q_0$ and $K_{all}$; the chosen values provide 
the best description of the
average multiplicity not only at LEP-1 $cms$ energy but also at lower $cms$
energies}. 
Table~\ref{tableparameter} shows a sample of  parton multiplicities 
obtained by ARIADNE for different parameters $Q_0$ and $\Lambda$. 
By comparing the parton results with the experimental data on all hadrons, 
one concludes that the best set of parameters is given by $Q_0$ = 0.2 GeV
and $\lambda$ = 0.015. It is remarkable to notice that the same value of
$\lambda$ has been obtained in previous studies on jet and particle
multiplicities\cite{lomult}, whereas the present  value of $Q_0$ 
is slightly smaller than in the previous case. This difference can
be related to the different approximation schemes applied.

\begin{table}     
 \begin{center}
 \vspace{4mm}
 \begin{tabular}{||c|c|c|c||}
 \hline 
 & & $\sqrt{s}$ = 91.2 GeV & $\sqrt{s}$ = 14 GeV \\ 
\hline 
 & & $\bar n_{all} = 31.0 \pm 1.1$ & $\bar n_{all} = 13.95\pm0.65$ \\ 
 \hline
  $Q_0$ & $\lambda$  & $\bar n_p$ &  $\bar n_p$ \\
  \hline
  0.6  (def.) & 1 (def.) & 8.21 &  4.2  \\ 
  0.27        & 0.1      & 19.2  & 9.7  \\ 
  0.27        & 0.01      & 24.8 & 12.5 \\ 
  0.23        & 0.01      & 29.4 & 13.4  \\ 
  0.2         & 0.015     & 30.2 & 13.8  \\ 
 \hline
 \end{tabular}
 \end{center}
\caption{Predictions by ARIADNE at parton level for the total average
multiplicity $\bar n_p$
in $e^+e^-$ annihilation at $\protect\sqrt{s}$ = 14 and 91.2 GeV 
for different values of the parameters $Q_0$ and $\protect\lambda$, compared 
with the experimental results for all hadrons $\bar n_{all}$ 
(taken as 3/2 the value for
charged hadrons).}
\label{tableparameter}
\end{table}

As a further test of LPHD, 
let us compare the predictions of ARIADNE using the new 
parameters with experimental data on factorial moments in one hemisphere
as measured by the OPAL Collaboration\cite{opalmom}. This is done in
Table~\ref{tablemoments}, which clearly gives further evidence in support
of LPHD using the Monte Carlo results at the parton level.
 In this comparison  we have assumed that
the normalized higher order global moments,
$F_q$, are equal for charged hadrons and for all hadrons.

After having fixed the set of parameters of ARIADNE which are
consistent with the picture of LPHD, let us now study the predictions 
 for the cut moments at parton level; the dependence of
the $k_\perp^{cut}$ and $k^{cut}$ moments on the cut parameter 
in one hemisphere defined via the thrust
axis is shown in
Fig.~\ref{figpart02}. According to our analytical 
calculations, we expect that the
$k_\perp$ moments approach unity as $\ln k_\perp/Q_0$ for 
$k_\perp \to Q_0$. The cut 
moments predicted by ARIADNE show indeed such behaviour for small values of
the cut, $k_\perp^{cut} \le$ 4 GeV; however, at very small $k_\perp^{cut} 
\to Q_0$, the moments
do not reach the predicted Poissonian value of 1, but saturate at a 
value larger than 1. Such effects are expected from 
threshold effects close to the infrared cutoff,  in particular, from the boost 
needed to relate the overall rest frame of the collision 
to the single dipole rest frame where each new emission takes place. 

The $k^{cut}$ moments in Fig.~\ref{figpart02}b rise to rather large values
at intermediate scales and  bend at small $k^{cut}$; 
the important point here is that they  reach 
finite values much above unity, showing that in spherically  symmetric 
phase space there is no 
Poissonian regime at very small momentum cuts. 
Both types of moments approach the same values for large $k_\perp^{cut} \sim
k^{cut}$ corresponding to the global moments as it should be. 

To investigate in more detail the behaviour of the $k_\perp$ moments in the
kinematic region close to $Q_0$, let us now examine a new simulation
with the same value of $\lambda$ but the larger value  $Q_0$ = 0.4 GeV. 
Results of this run are shown in Fig.~\ref{figpart} for both
$k_\perp^{cut}$ and
$k^{cut}$ moments. In this case, in which the cascade is stopped earlier, the
situation is clearer: the $k_\perp$ moments approach the Poissonian
value one for $k_\perp^{cut} \to Q_0$ as theoretically expected, and stay there 
below the cutoff. The $k^{cut}$ moments saturate, on the other hand,  
at values above one  as before.

\begin{table}     
 \begin{center}
 \vspace{4mm}
 \begin{tabular}{||c|c|c|c|c||}
 \hline
  & $\bar n$ & $F^{(2)}$ & $F^{(3)}$ & $F^{(4)}$  \\
 \hline
 Partons  & 30.18 & 1.09 & 1.30 & 1.68 \\
 \hline
 Hadrons & $30.3 \pm 0.6$ & $1.0820\pm 0.0052$ &
$1.275\pm0.019$ & $1.637\pm0.047$ \\
\hline
 \end{tabular}
 \end{center}
\caption{The mean multiplicity $\bar n$ ($\bar n_p$ or $\bar n_{all}$) 
and the first three normalized factorial moments 
in one hemisphere defined through the thrust axis predicted by 
ARIADNE at parton level with parameters $Q_0$ = 0.2 GeV and $\lambda$ = 0.015 
and experimentally measured by the OPAL Collaboration\protect\cite{opalmom}. 
Errors on data are the sum of statistical and systematic errors.}
\label{tablemoments}
\end{table}

From these results we conclude that the more complete Monte Carlo
calculation confirms the essential features of our analytical calculation in
the region of small cut variables: for the $k_\perp^{cut}$-cut the moments
vanish linearly as $\ln (k_\perp^{cut}/Q_0)$, whereas for the $k^{cut}$-cut
the moments approach a constant value. However, the slope and the constant
values are different from the analytical predictions and also the
dependence for large values of the cut momentum. 
We attribute these differences to our simplified DLA calculations which does
not take into account energy conservation constraints and the nonleading
parts of the parton splitting functions.

Finally, we show in Fig.~\ref{fighadr} the predictions for the cut moments 
 both with
cylindrical and spherical cuts obtained at hadron level by running  ARIADNE 
with default parameters for $e^+e^-$ annihilations 
at $cms$ energy of LEP-1. Data refer again to particles in
one hemisphere, defined through the thrust-axis. Electrons and
positrons have been subtracted to avoid the lepton contaminations of the 
hadronic signal; 
by keeping the electron pairs, a peak at very low values
of the cut momentum would appear, masking the eventual physical signal. This
result is consistent with an old observation\cite{int} done within 
the study of intermittency phenomena that $e^+e^-$ pairs from $\pi^0$ decays 
strongly affect the experimental results in very short-range correlations. 

As expected, for large values of the cut, the two  moments go to a 
 common value corresponding to 
the global moments of the multiplicity distribution without any cut; 
the obtained results are close but up to about 10\%  smaller than 
the experimental values for the
global moments directly measured by the OPAL Collaboration\cite{opalmom}.   	
For small values of the cut, the two moments show a different behavior; 
the cylindrical symmetric cut moments for $k_\perp^{cut} \leq 5$ GeV first
tend towards unity, in agreement
with our theoretical expectations; however, they rise again 
for very small values of the cut. This effect does not occur at the parton
level and therefore has to be associated  with  hadronization
corrections, in particular production and decay of 
resonances\footnote{In order to check this point, we have swichted off the
perturbative cascade within ARIADNE and found that all moments approach
unity for $k_\perp^{cut} >$  1 GeV, while the peaks at small values of
$k_\perp^{cut}$ remain. If resonances are not allowed to decay, all moments
become more reduced at small $k_\perp^{cut}$ , but small peaks 
are still visible.}. 
The spherical symmetric cut moments show a  step around a few GeV's
which separates two plateaus. The depletion at large $k^{cut}$ 
 could be interpreted again as 
the result of energy-momentum constraints not included in our DLA
calculations. It is remarkable to notice that the moments saturate and do
not fall down for small values of the cut momentum, very much like in our
analytical calculation. However, in view of the parton level results for the
same quantity we consider this coincidence as accidental. 

By comparing the partonic and the hadronic predictions, one sees  
that the LPHD is approximately confirmed within ARIADNE 
at the level of global moments and for large values of the cut momenta 
(right hand side of the plots) but is clearly
violated for soft particles (left hand side of the plots). So we conclude
that the parton level predictions show
indeed the peculiar features 
predicted analytically within the perturbative approach, namely the near
Poissonian production of soft particles as consequence of coherent multigluon
emission, 
whereas the same features are softened and partly washed out in the model
with  string fragmentation. 
It will be interesting to find out the experimental situation. 

\section{Conclusions}
We suggest analysing multiparticle production in  restricted
phase space regions with variable cut $k^{cut}_\perp$ in 
 the transverse momenta of the particles, 
and, for comparison, with variable momentum cut $k^{cut}$. 
We have derived the evolution equations for the
multiplicity moments within the DLA of perturbative QCD, from which
their dependence on the jet virtuality and on both cut variables can be
obtained.
Results for moments up to order
$q=4$ are given explicitly in case of fixed coupling 
by the first terms of the perturbative
expansion.

The normalized factorial moments are found to
approach a linear behaviour for small values of the transverse momentum cut
with $F^{(q)}\to 1$
for $k^{cut}_\perp \to Q_0$
corresponding to a Poisson distribution, as the parton showering is
uniformly suppressed at small $k_\perp$ for any rapidity. On the
other hand, for the momentum cut, this is not the case and the
moments approach a finite value for the minimal value of $k^{cut}$.

The DLA takes into account only the leading singularities but respects
the soft gluon coherence  which is implemented through the
angular ordering prescription. We have shown that this property is
responsible for obtaining a Poisson distribution in the soft limit.
As a check of our analytical results, we have compared them with 
the results of a parton Monte Carlo program 
which fully takes into account the constraints from energy-momentum
conservation and the complete parton splitting functions. The factorial
moments for decreasing $k^{cut}_\perp$ continuously decrease to values
close to unity, contrary to the case of the momentum cut, as expected
from the analytical calculations.
At the quantitative level, however, 
the DLA results show considerable deviations
from those of the more realistic Monte Carlo calculations, as already
observed in the case of global moments.

It will be interesting to find out to what extent the experimental
data follow the perturbative results as one might expect from previous
successes
of the LPHD picture. Final state interactions, in particular resonance
production, could severely disturb the perturbative results, especially
in the soft region, as it is suggested indeed by 
the results of a Monte Carlo program  which implements a model of the 
hadronization phase. By the proposed studies 
the limitations of the perturbative
predictions could be explored in genuine multiparticle correlations.

\section*{Acknowledgements}

JW thanks the Theory Group at the Werner-Heisenberg Institute
of Physics for the hospitality and the financial support.
This work is supported in part by the Polish Committee
for Scientific Research under grants No 2P03B19609 and 2P03B04412.

\newpage
\section*{Appendix A}
We derive here the evolution equation for the second cumulant in the spherically cut phase space following 
in detail the steps outlined in Section 4.1 . Integrating Eqs.(\ref{gamman},\ref{dnest}) for $q=2$
 gives
\begin{eqnarray}\lefteqn{c^{(2)}(\kc,P,\Theta)=2\int_{k_1<\kc} d^3k_1{\cal M}_P(k_1)\int_{k_2<\kc}
 d^3k_2\rho^{(1)}_{k_1}(k_2) } \nonumber \\
 &&+ \int_{\Gamma_{I+II+III}} d^3K{\cal M}_P(K)\int_{k_1<\kc} d^3k_1\int_{k_2<\kc}
 d^3k_2\rho^{(2)}_K(k_1,k_2),  \label{c2abs} 
\end{eqnarray}
where the order of integrals in the second term was interchanged. The full phase space of the 
intermediate parent $\vec{K}$ is displayed in Fig.\ref{fkin1}. As in the one-parton case, only  
regions I and II contribute since both momenta of final partons are limited from above by the 
cut-off $\kc$. This in turn restricts from below their emission angles and consequently, due to 
the angular ordering, the emission angle of the parent $K$. We then split explicitly the
 remaining integral into regions $\Gamma_I$ and $\Gamma_{II}$. Simplifying also the 
first term gives
\begin{eqnarray}
\lefteqn{c^{(2)}(\kc,P,\Theta)=2\int_{Q_0/\kc} {d\vartheta_1\over\vartheta_1}
\int_{Q_0/\vartheta_1} {dk_1\over k_1}\ad(k_1\vartheta_1) \bar{n}(k_1\vartheta_1) }
\nonumber \\&& \nonumber 
\lefteqn{ + \int_{Q_0/\kc}^{\Theta}{d\vartheta\over\vartheta}
\int_{Q_0/\vartheta}^{\kc} {dK\over K} \ad(K\vartheta) f^{(2)}(K\vartheta)}\\&&+ 
\int_{Q_0/\kc}^{\Theta}{d\vartheta\over\vartheta}\int_{\kc}^{P} {dK\over K} 
\ad(K\vartheta) f^{(2)}_c(\kc,K,\vartheta). 
\end{eqnarray}
In the first term, the $k_2$ integral gave just the global average multiplicity at 
the scale $k_1\vartheta$ since the restriction $k_2<\kc$ is weaker than $k_2<k_1<\kc$ which
 is satisfied for this subprocess anyway. Similarly in the second term, momentum of the 
parent $K$ in the region $\Gamma_I$ is smaller than the cut-off, hence the phase space of 
the final partons is not restrited by $\kc$ but by the momentum of the parent $K<\kc$. 
Therefore the global moment at the scale $K\vartheta$ results. On the other hand in the third 
term a parent $K$ is harder than the cut-off and phase space of the final partons is indeed 
restricted by $\kc$. Consequently integrating over $k_1$ and $k_2$ gives the cut moment. It is 
natural to introduce the logarithmic variables $Y=\ln(P\Theta/Q_0)\;X=\ln(\kc\Theta/Q_0)$. 
Then $c^{(2)}(\kc,P,\Theta)=c^{(2)}(X,Y)$. Finally we observe that, due to the evolution equation
 for the {\em global} moments, Eq.(\ref{fcmom}), the first two terms combine into the 
global cumlant at the scale $\kc\Theta$, hence
\begin{equation}c^{(2)}(X,Y)=c^{(2)}(X)+\int_0^X dx \int_x^{Y-X+x} dy\ad(y)f^{(2)}(x,y).
\end{equation}
\section*{Appendix B}

In this Appendix we collect a sample of higher order expressions for the cut
 moments. The perturbative expansion for a generic (spherically or cylindrically 
cut) moment reads
\begin{equation}
f_{k_{max}}^{(q)}(Z,Y)=\sum_{k=q}^{k_{max}} \ad^k f^{(q)}_k(Z,Y),
\end{equation}
where the cut-off $Z=X$ for moments in a spherical phase 
space and $Z=X_{\perp}$ for a cylindrical cut. Our results for the 
coefficients $f^{(q)}_{sph,k}(X,Y)$ and $f^{(q)}_{cyl,k}(X_{\perp},Y)$ are 
shown in Table 3 and Table 4 respectively. The formulas for the spherically 
cut moments were generated (by Mathematica) recursively in $k$ and in $q$ 
from the evolution equation, Eq.(\ref{cqmom}) coupled with Eq.(\ref{fcrec}). 
The formulas for cylindrically cut moments were obtained by integrating 
spherically cut moments according to Eq.(\ref{cqrel}). As one of the 
consistency tests we have checked that the
 results obtained from Tables 3 and 4 for the normalized moments 
$F_{k_{max}}^{(q)} =  f_{k_{max}}^{(q)}/ (f_{k_{max}}^{(1)})^q$, 
in the leading order of $X$ and $X_\perp/Y$, agree
with those from Eqs. (\ref{Kk}) and (\ref{FqLO}) respectively.

\newpage

  \begin{table}[t]    
  \begin{center}
   \begin{tabular}{|c|c|cccccccc|} \hline\hline
   {\em q} & {\em k} & \multicolumn{8}{c|}{ $f^{(q)}_{sph,k}(X,Y)$ }  \\
   \hline
   & 1 & \multicolumn{8}{l|}{$\frac{{X^2}}{2} $} \\
 1 & 2 & \multicolumn{8}{l|}{$-\frac{{X^4}}{8}+\frac{ {X^3} Y}{6} $}          \\
   & 3 & \multicolumn{8}{l|}{$\frac{{X^6}}{72}-\frac{ {X^5} Y}{30}
                                 +\frac{ {X^4} {Y^2}}{48} $}           \\
\hline
   & 2 & \multicolumn{8}{l|}{$\frac{{X^4}}{3} $} \\
 2 & 3 & \multicolumn{8}{l|}{$ \frac{{X^6}}{18}+\frac{7 {X^5} Y}{30} $}          \\
   & 4 & \multicolumn{8}{l|}{$\frac{{X^8}}{240}+\frac{19 {X^7} Y}{630}
                                 +\frac{49 {X^6} {Y^2}}{720} $}           \\
\hline
   & 3 & \multicolumn{8}{l|}{$ \frac{17 {X^6}}{60}$} \\
 3 & 4 & \multicolumn{8}{l|}{$\frac{79 {X^8}}{1120}+\frac{43 {X^7} Y}{140} $}          \\
   & 5 & \multicolumn{8}{l|}{$\frac{23 {X^{10}}}{2800}+\frac{647 {X^9} Y}{10080}+
\frac{323 {X^8} {Y^2}}{2240} $}           \\
\hline
   & 4 & \multicolumn{8}{l|}{$ \frac{31 {X^8}}{105}$} \\
 4 & 5 & \multicolumn{8}{l|}{$ \frac{737 {X^{10}}}{7560}+\frac{412 {X^9} Y}{945}$}          \\
   & 6 & \multicolumn{8}{l|}{$\frac{76603 {X^{12}}}{4989600}+\frac{4757 {X^{11}} Y}{37800}+
\frac{6161 {X^{10}} {Y^2}}{21600} $}           \\
   \hline\hline  
   \end{tabular}
  \end{center}
\caption{Perturbative expansion of the spherically cut moments}
   \end{table}

\vspace{5.0cm} 
\phantom{ciao}
\mbox{}

\newpage

 \begin{table}[t]
  \begin{center}
   \begin{tabular}{|c|c|cccccccc|} \hline\hline
   {\em q} & {\em k} & \multicolumn{8}{c|}{ $f^{(q)}_{cyl,k}(\xt,Y)$ }  \\
   \hline
   & 1 & \multicolumn{8}{l|}{$ -\frac{ \xt^2 }{2}+{\xt} {Y}$} \\
 1 & 2 & \multicolumn{8}{l|}{$-\frac{{\xt^4}}{8}-\frac{{\xt^3} Y}{3}+
                    \frac{ {\xt^2} {Y^2}}{4} $}          \\
   & 3 & \multicolumn{8}{l|}{$-\frac{{\xt^6}}{72}+\frac{ {\xt^5} Y}{20}
        -\frac{ {\xt^4} Y^2}{16}
+\frac{ {\xt^3} {Y^3}}{36}  $}           \\
\hline
   & 2 & \multicolumn{8}{l|}{$ -\frac{2 {\xt^3} Y}{3}+{\xt^2} {Y^2}$} \\
 2 & 3 & \multicolumn{8}{l|}{$-\frac{{\xt^6}}{36}+\frac{{\xt^5} Y}{3}-
                    \frac{3 {\xt^4} {Y^2}}{4}+\frac{{\xt^3} {Y^3}}{2} $}          \\
   & 4 & \multicolumn{8}{l|}{$\frac{{\xt^8}}{90}-\frac{11 {\xt^7} Y}{120}
        +\frac{89 {\xt^6} Y^2}{360}
-\frac{101 {\xt^5} {Y^3}}{360}+\frac{17 {\xt^4} {Y^4}}{144} $}           \\
\hline
   & 3 & \multicolumn{8}{l|}{$\frac{{\xt^6}}{12}-\frac{3 {\xt^5} Y}{10}-
\frac{{\xt^4} {Y^2}}{2}+{\xt^3} {Y^3} $} \\
 3 & 4 & \multicolumn{8}{l|}{$-\frac{29 {\xt^8}}{480}+\frac{83 {\xt^7} Y}{420}
+\frac{11 {\xt^6}{Y^2}}{60}-{\xt^5} {Y^3}+\frac{3 {\xt^4} {Y^4}}{4} $}          \\
   & 5 & \multicolumn{8}{l|}{$\frac{551 {\xt^{10}}}{33600}
-\frac{589 {\xt^9} Y}{10080}
-\frac{97 {\xt^8} {Y^2}}{2240}+\frac{117 {\xt^7} {Y^3}}{280}
-\frac{857 {\xt^6} {Y^4}}{1440}+\frac{13 {\xt^5} {Y^5}}{48}$}           \\
\hline
   & 4 & \multicolumn{8}{l|}{$\frac{{\xt^8}}{15}+\frac{2 {\xt^7} Y}{21}-
\frac{13 {\xt^6} {Y^2}}{15}+{\xt^4} {Y^4} $} \\
 4 & 5 & \multicolumn{8}{l|}{$-\frac{11 {\xt^{10}}}{280}
-\frac{383 {\xt^9} Y}{3780}
+\frac{169 {\xt^8} {Y^2}}{210}-\frac{11 {\xt^7} {Y^3}}{15}
-\frac{5 {\xt^6} {Y^4}}{6}+{\xt^5} {Y^5} $}          \\
   & 6 & \multicolumn{8}{l|}{$
\frac{1087 {\xt^{12}}}      {151200} + \frac{14941 {\xt^{11}} Y} {277200}
-\frac{683 {\xt^{10}} {Y^2}}{1800}   + \frac{27017 {\xt^9}{Y^3}} {45360}
+\frac{187 {\xt^8} {Y^4}}   {2520}-  
 \frac{37 {\xt^7} {Y^5}}    {45}  +    \frac{35 {\xt^6} {Y^6}}   {72}
                              $}      \\     
   \hline\hline  
   \end{tabular}
  \end{center}
\caption{Perturbative expansion of the cylindrically cut moments}
   \end{table}      

\vspace{6.0cm}
\mbox{}
\mbox{}
\phantom{ciao}
\mbox{}
\phantom{ciao}

\newpage
\section*{Appendix C}
We will analyse in detail the steps leading to the evolution equations for the cylindrically cut moments, 
Eqs.(\ref{navrel},\ref{cqrel}).    The cut multiplicity is defined, c.f. Fig.~\ref{kin3}
\begin{eqnarray}
\lefteqn{
  \bar{n}(\kct,P,\Theta)=\int_{Q_0/P}^{\Theta}d\vartheta  
\int_{Q_0/\vartheta,\;k\vartheta<\kct}^P d k \rho(\vartheta,k,P) } \nonumber 
\\&&=
\int_{Q_0/P}^{\kct/P}d\vartheta \int_{Q_0/\vartheta}^P d k \rho(\vartheta,k,P)
+\int_{\kct/P}^{\Theta}d\vartheta \int_{Q_0/\vartheta}^{\kct/\vartheta} d k
\rho(\vartheta,k,P).
\labl{ad1}
\end{eqnarray}
The first term (region A) is just the global multiplicity at the scale $\kct$. 
In the second term we use Eq.(\ref{tkk}).
\begin{eqnarray}
\nonumber 
\lefteqn{\bar{n}(\kct,P,\Theta)=\bar{n}(\kct)+ \int_{\kct/P}^{\Theta}{d\vartheta\over\vartheta} 
\int_{Q_0/\vartheta}^{\kct/\vartheta} d k }\\&&
\left\{{\ad(k\vartheta)\over k}+\int_k^P{dK\over K}\ad(K\vartheta)
\int_{Q_0/k}^{\vartheta}d\Theta_{Kk}\rho(\Theta_{Kk},k,K)\right\}. 
\end{eqnarray} 
which after changing the orders of $dk$ and $dK$ integrals gives 
\begin{eqnarray}
\nonumber \lefteqn{\bar{n}(\kct,P,\Theta)=\bar{n}(\kct)+\int_{\kct/P}^{\Theta}{d\vartheta\over\vartheta} 
\int_{Q_0/\vartheta}^{\kct/\vartheta} {d k\over k} \ad(k\vartheta)+}\\&&
\int_{\kct/P}^{\Theta}{d\vartheta\over\vartheta} \int_{Q_0/\vartheta}^{\kct/\vartheta}{dK\over K} \ad(K\vartheta) 
\int_{Q_0/\vartheta}^K dk \int_{Q_0/k}^{\vartheta}d\Theta_{Kk}\rho(\Theta_{Kk},k,K)+ 
\nonumber 
\\&&
\int_{\kct/P}^{\Theta}{d\vartheta\over\vartheta} \int_{\kct/\vartheta}^P{dK\over K} \ad(K\vartheta) 
\int_{Q_0/\vartheta}^{\kct/\vartheta} dk
\int_{Q_0/k}^{\vartheta}d\Theta_{Kk}\rho(\Theta_{Kk},k,K). 
\end{eqnarray}
In the third term integrations over $k$ and $\Theta_{Kk}$ extend over the full phase space of the $K$ jet, 
hence they give global multiplicity. In the fourth term the child momentum is
restricted, therefore the 
momentum cut multiplicity results 
\begin{eqnarray}
\nonumber 
\lefteqn{\bar{n}(\kct,P,\Theta)=\bar{n}(\kct)+\int_{\kct/P}^{\Theta}{d\vartheta\over\vartheta} 
\int_{Q_0/\vartheta}^{\kct/\vartheta} {d k\over k} \ad(k\vartheta)+}\\
\nonumber && \lefteqn{
\int_{\kct/P}^{\Theta}{d\vartheta\over\vartheta} \int_{Q_0/\vartheta}^{\kct/\vartheta}{dK\over K} 
\ad(K\vartheta) \bar{n}(K\vartheta) +}\\&&\int_{\kct/P}^{\Theta}{d\vartheta\over\vartheta} 
\int_{\kct/\vartheta}^P{dK\over K} \ad(K\vartheta) \bar{n}_{sph}(\kct/\vartheta,K,\vartheta)
\end{eqnarray} 
which when transformed to the logarithmic variables gives Eq.(\ref{navrel}). 

Let us now study the second cumulant, 
Eqs.(\ref{phsp2}). Inserting Eq.(\ref{g2diff}) in (\ref{c2split}) gives
\begin{eqnarray}
\nonumber \lefteqn{c_{cyl}^{(2)}(\kct,P,\Theta)=c^{(2)}(\kct)+}\\ 
\lefteqn{2\int_{\kct/P}^{\Theta}{d\vartheta_1\over\vartheta_1}
\int_{Q_0/\vartheta_1}^{\kct/\vartheta_1} {dk_1\over k_1} \ad(k_1\vartheta_1)}&&
\nonumber \\
\nonumber &&
\lefteqn{
\int_{ Q_0 /k_1}^{\vartheta_{1}} d\vartheta_{12} 
\int_{Q_0/\vartheta_{12}}^{k_1} dk_2 \rho^{(1)}(k_2,\vartheta_{12},k_1)+}\\
\lefteqn{2\int_{\kct/P}^{\Theta}{d\vartheta_1\over\vartheta_1}
\int_{Q_0/\vartheta_1}^{\kct/\vartheta_1} \frac{dK}{K}\ad(K\vartheta_1)
\int_{ Q_0 /K}^{\vartheta_{1}} d\vartheta_{12}}&& \nonumber \\ \nonumber && \lefteqn{
\int_{Q_0/\vartheta_{12}}^K dk_1\int_{Q_0/k_1}^{\vartheta_{12}}d\Theta_{Kk_1}\rho^{(1)}(\Theta_{Kk_1},k_1,K)}\\ 
\nonumber &&
\lefteqn{
\int_{Q_0/\vartheta_{12}}^K dk_2\rho^{(1)}(\vartheta_{12},k_2,K)+}\\
\nonumber 
\lefteqn{2\int_{\kct/P}^{\Theta}{d\vartheta_1\over\vartheta_1}
\int_{\kct/\vartheta_1}^P \frac{dK}{K}\ad(K\vartheta_1)
\int_{ Q_0 \vartheta_1/\kct}^{\vartheta_{1}} d\vartheta_{12}}&&\\ \nonumber &&
\lefteqn{
\int_{Q_0/\vartheta_{12}}^{\kct/\vartheta_1} dk_1
\int_{Q_0/k_1}^{\vartheta_{12}}d\Theta_{Kk_1}\rho^{(1)}(\Theta_{Kk_1},k_1,K)}\\
\nonumber && \lefteqn{
\int_{Q_0/\vartheta_{12}}^{\kct/\vartheta_1} dk_2\rho^{(1)}(\vartheta_{12},k_2,K)+}\\
\lefteqn{2\int_{\kct/P}^{\Theta}{d\vartheta_1\over\vartheta_1}
\int_{Q_0/\vartheta_1}^{\kct/\vartheta_1} \frac{dK}{K}\ad(K\vartheta_1)
\int_{ Q_0 /K}^{\vartheta_{1}} d\vartheta_{12}}&& \nonumber \\&& \nonumber \lefteqn{
\int_{Q_0/\vartheta_{12}}^K dk_1
\int_{Q_0/\vartheta_{12}}^K dk_2
\int_{\vartheta_{12}}^{\vartheta_1} d\Theta_{Kk_1}
    \Gamma^{(2)}( \Theta_{Kk_1}, \vartheta_{12},k_1,k_2,K)+}\\
\nonumber \lefteqn{
2\int_{\kct/P}^{\Theta}{d\vartheta_1\over\vartheta_1}
\int_{\kct/\vartheta_1}^P \frac{dK}{K}\ad(K\vartheta_1)
\int_{ Q_0 \vartheta_1/\kct}^{\vartheta_{1}} d\vartheta_{12}}&&\\&&
\int_{Q_0/\vartheta_{12}}^{\kct/\vartheta_1} dk_1
\int_{Q_0/\vartheta_{12}}^{\kct/\vartheta_1} dk_2
\int_{\vartheta_{12}}^{\vartheta_{1}} d\Theta_{Kk_1} 
    \Gamma^{(2)}( \Theta_{Kk_1}, \vartheta_{12},k_1,k_2,K).  
\end{eqnarray}
In the second term the integrals over $\vartheta_{12}$ and $k_1$have been interchanged. 
Since this term describes the process $P\rightarrow k_1\rightarrow k_2$, a momentum $k_2$
 is limited in fact by
 $k_1<\kct/\vartheta_1$ and not by the cut-off $\kct/\vartheta_1$ itself.
In the remaining four terms the $K$ integrations have been pulled in front of the $k_1,k_2$ and $\vartheta_{12}$ 
integrations. It is then convenient to split the full range of the intermediate parent momentum $K$ according to
 the cut-off $\kct/\vartheta_1$, similarly to the average multiplicity (q=1) case discussed above. 
For $K$ below the cut-off the inner integrals cover the full phase space of the $K$ jet, while for $K$ 
above the cut-off only the spherically restricted integrals over children momenta $k_1$ and $k_2$ occur.
 All these inner integrals can be cast into the global or spherically restricted moments. We have
\begin{eqnarray}
\lefteqn{c_{cyl}^{(2)}(\kct,P,\Theta)=c^{(2)}(\kct)+} \nonumber \\&&
\lefteqn{2\int_{\kct/P}^{\Theta}{d\vartheta_1\over\vartheta_1}
\int_{Q_0/\vartheta_1}^{\kct/\vartheta_1} {dk_1\over k_1}\ad(k_1\vartheta_1)
       \bar{n}(k_1\vartheta_1)+}
\nonumber \\ \nonumber &&
\lefteqn{\int_{\kct/P}^{\Theta}{d\vartheta_1\over\vartheta_1}
\int_{Q_0/\vartheta_1}^{\kct/\vartheta_1} \frac{dK}{K}\ad(K\vartheta_1)\bar{n}^2(K\vartheta_1)+}
\\&&
\lefteqn{\int_{\kct/P}^{\Theta}{d\vartheta_1\over\vartheta_1}\int_{\kct/\vartheta_1}^P \frac{dK}{K}
\ad(K\vartheta_1)\bar{n}^2_{sph}(\kct/\vartheta_1,K,\vartheta_1)+} \nonumber \\&&
\lefteqn{\int_{\kct/P}^{\Theta}{d\vartheta_1\over\vartheta_1}
\int_{Q_0/\vartheta_1}^{\kct/\vartheta_1} \frac{dK}{K}\ad(K\vartheta_1)c^{(2)}(K\vartheta_1)+}
\nonumber \\&&
\int_{\kct/P}^{\Theta}{d\vartheta_1\over\vartheta_1}
\int_{\kct/\vartheta_1}^P \frac{dK}{K}\ad(K\vartheta_1)
   c^{(2)}_{sph}(\kct/\vartheta_1,K,\vartheta_1).
\end{eqnarray}
 Finally, introducing logarithmic variables, and after combining cumulants and squares of the average 
multiplicity into factorial moments, we obtain the $q=2$ counterpart of Eq.(\ref{cqrel}).


\newpage
\section*{Figure Captions}

\begin{figure}[p]
 \begin{center}
\mbox{\mbox{\epsfig{file=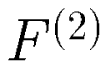,bbllx=5.cm,bblly=20cm,bburx=5.2cm,bbury=24.cm,width=0.2cm}}
\mbox{\epsfig{file=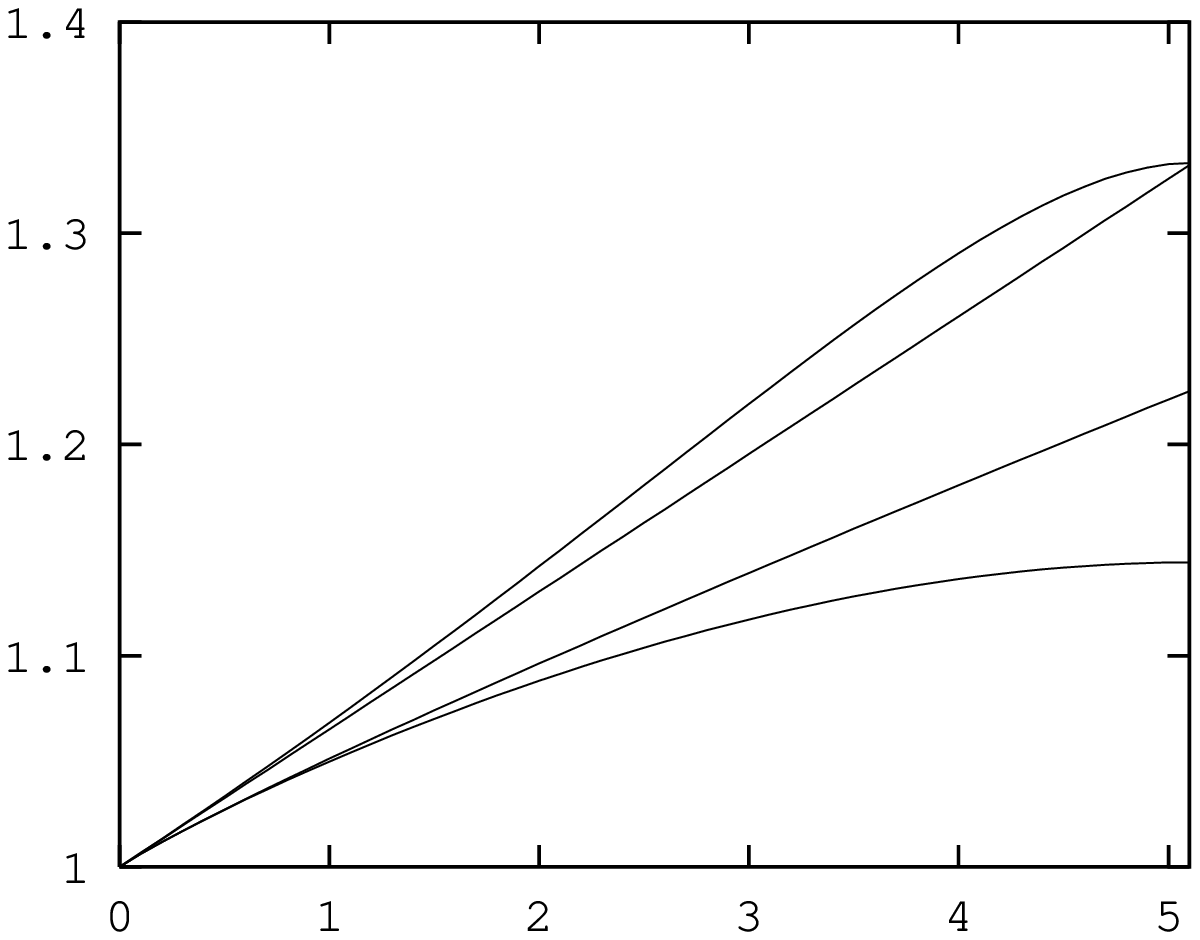,bbllx=3.cm,bblly=3.cm,
                     bburx=17cm,bbury=13.cm,width=12cm}}
 }         \end{center}
\vspace{-0.4cm}
\hspace{7.0cm} \large $X_\perp$
\caption{}
\label{figtheory1}
\end{figure}

\noindent  
{\bf Fig.~\ref{figtheory1}}.
 The normalized factorial moment $F^{(2)} = 1 + K^{(2)}$ for different
approximations in leading order. The lowest curve represents the
result~\eref{c2runful} for running $\alpha_s$, the next one the small
$X_\perp$ approximation~\eref{c2run}. The upper two curves represent the
fixed $\alpha_s$ result following from ~\eref{KqLO} and the linear
approximation~\eref{FqLO}. In all cases $Y$ = 5.1 and $\lambda$ = 0.015.

\begin{figure}[p]
 \begin{center}
\mbox{\epsfig{file=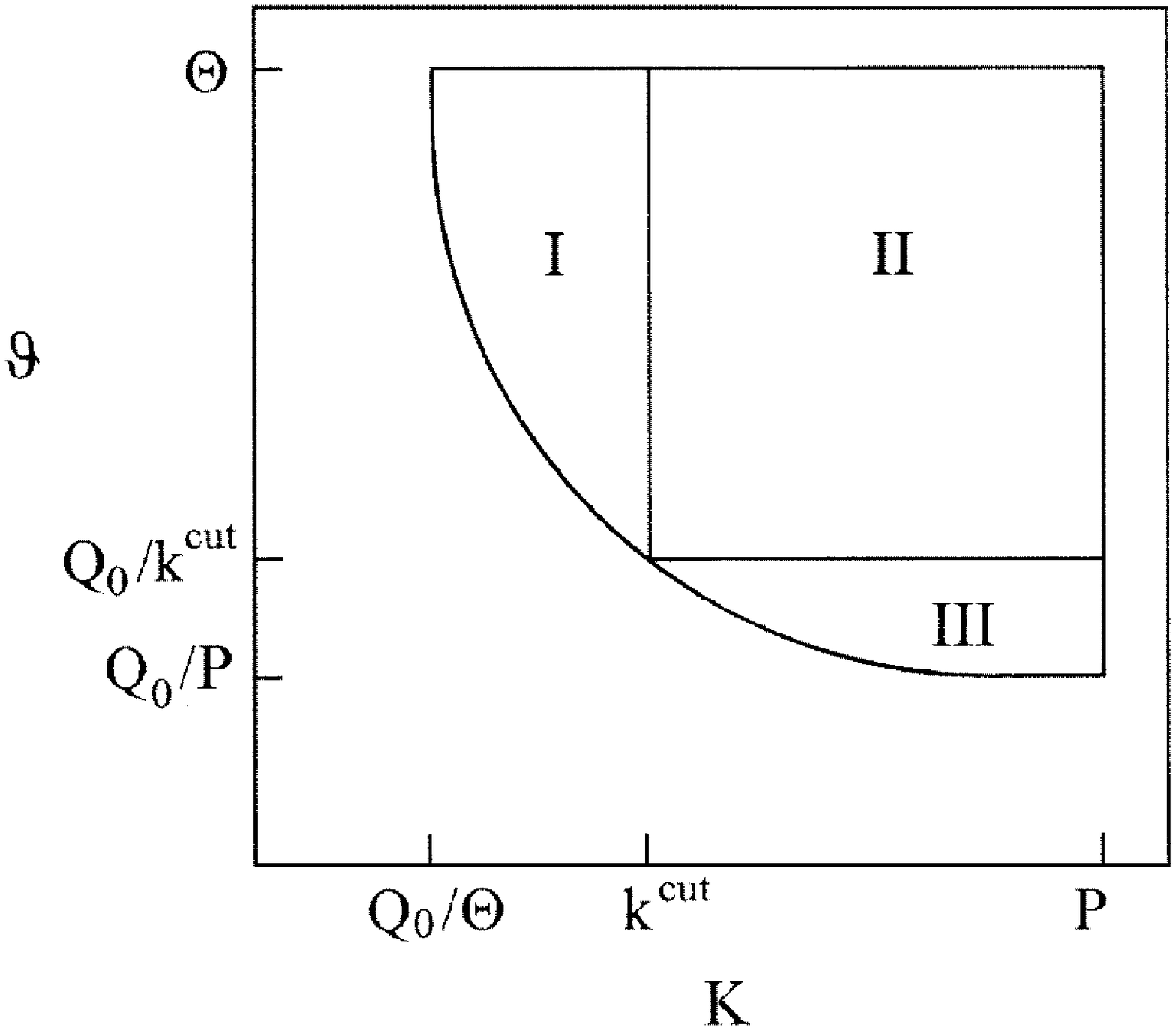,bbllx=-1.cm,bblly=8.5cm,bburx=24.cm,bbury=21.cm,width=12cm}}
          \end{center}
\caption{}
\label{fkin1}
\end{figure}

\noindent
{\bf Fig.~\ref{fkin1} }
  Phase space of an intermediate parent for the momentum cut moments,
c.f. Eqs.(\eref{nmom},\eref{c2abs}). Parents in region I generate
full jet with virtuality $\kc\Theta$, the ones in the region II contribute
to the cut moments, while region III does not contribute due the angular
ordering.

\begin{figure}[ht]
 \begin{center}
\mbox{\epsfig{file=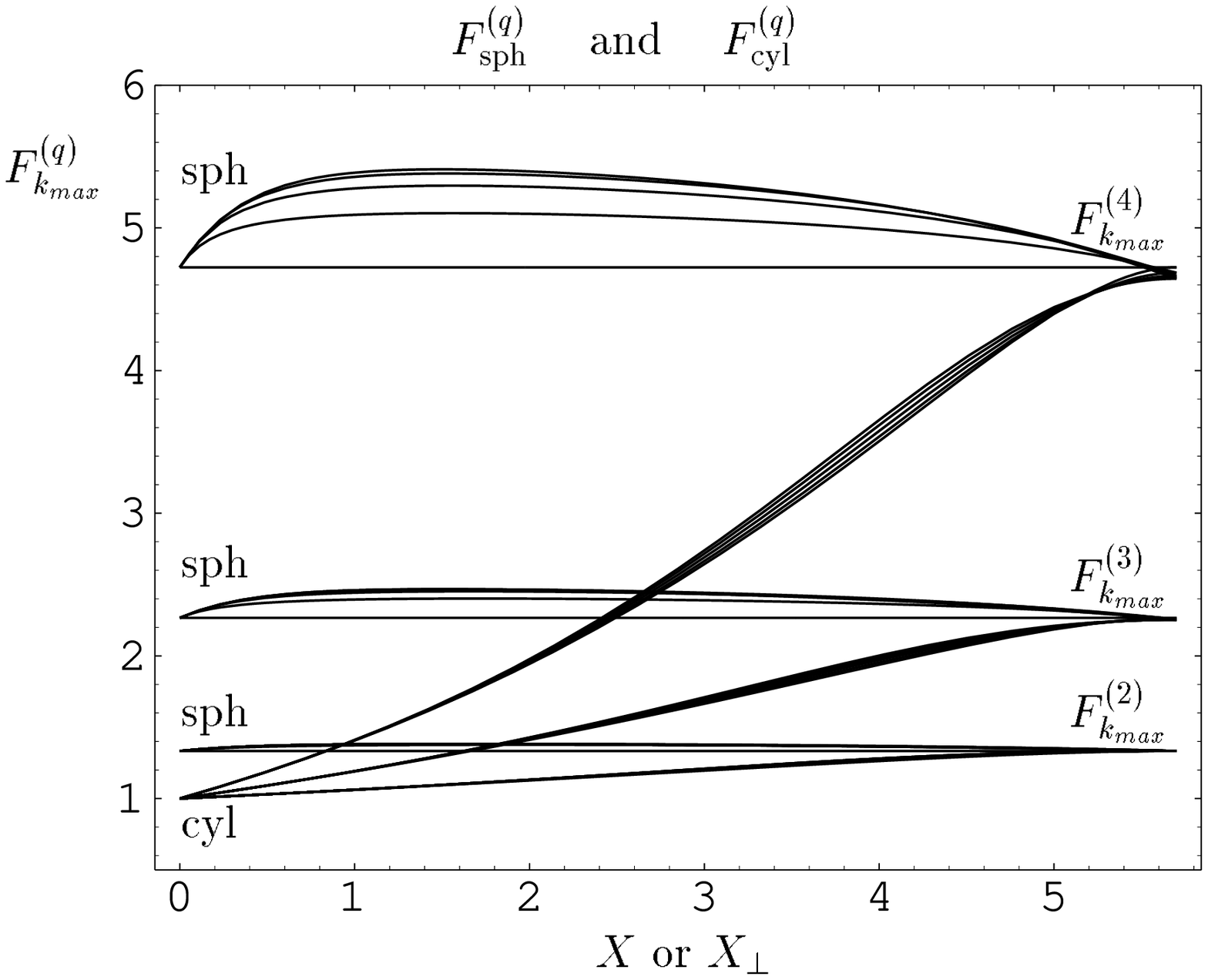,bbllx=3.cm,bblly=0.5cm,bburx=24.cm,bbury=18.5cm,width=18cm}}
          \end{center}
\caption{} 
\label{figtheory} 
\end{figure}

\noindent
{\bf Fig.~\ref{figtheory}}. 
DLA predictions for the cut-off dependence of the first
 three normalized moments $F^{(q)}_{k_{max}},\;q=2,3,4$ for $Y=5.7$.
 Both families, i.e. spherically (sph) and cyllindrically (cyl) cut moments,
 are shown.
 They coincide in the global limit (X or $X_{\perp} = Y$), but have distinctly
different threshold behaviour. Different lines which describe one moment 
correspond to different order of the perturbation theory included 
$k_{max},\;\;k_{max}=q,q+1,...,8$.

\begin{figure}[t]
 \begin{center}
\mbox{\epsfig{file=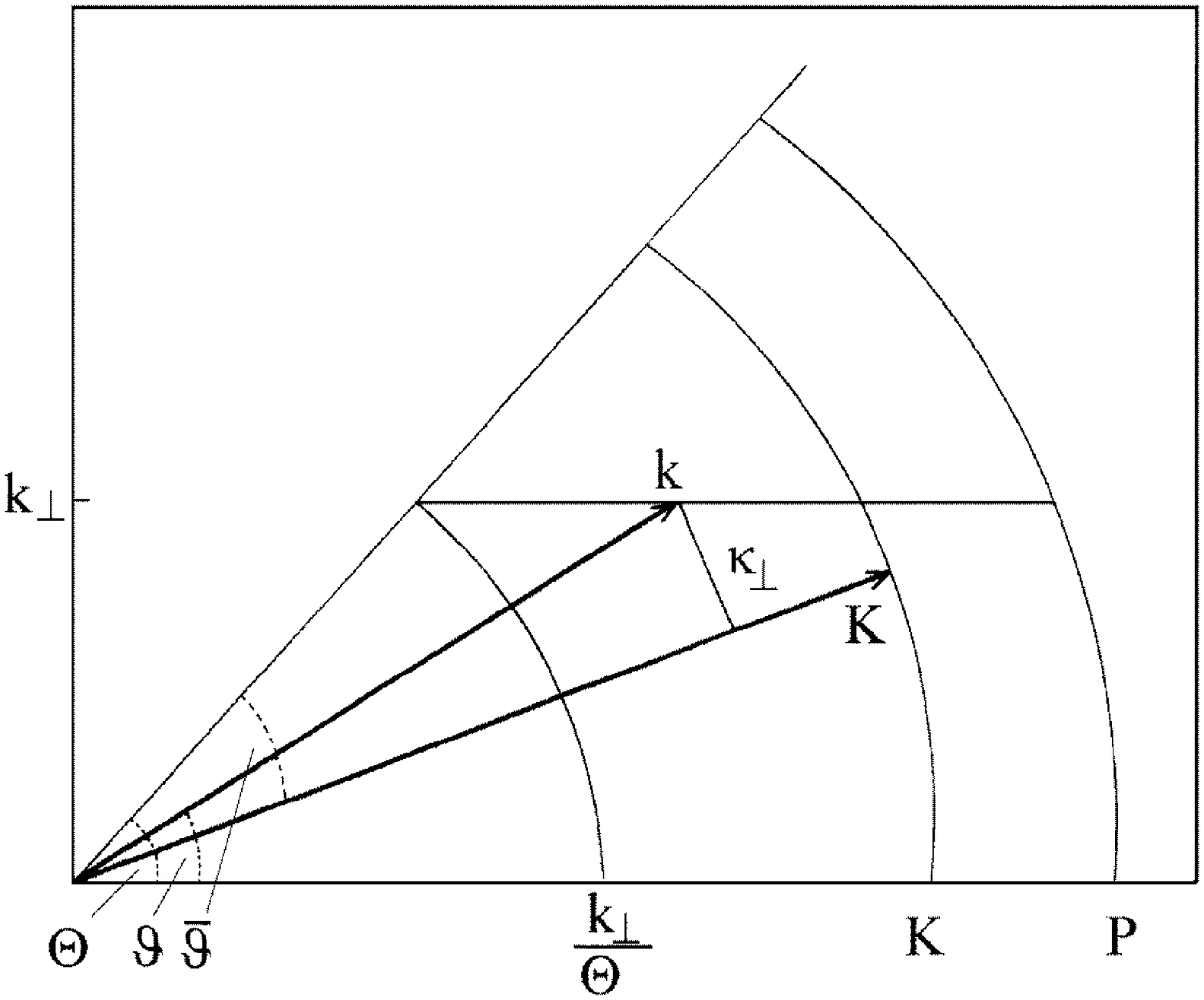,bbllx=-3.cm,bblly=3.cm,bburx=26.cm,bbury=19.cm,width=12cm}}
          \end{center}
\caption{}
\label{fkin2}
\end{figure}

\noindent 
{\bf Fig.~\ref{fkin2} }
Kinematics relevant for the evolution equations at fixed transverse
momentum, Eqs.(\eref{rhokt},\eref{navkt});  
$\bar{\vartheta}=\Theta\kappa_{\perp}/k_{\perp}$ is an angle between
a parent $\vec{K}$ and a parton emitted at the maximal angle $\Theta$.

\begin{figure}[b]
 \begin{center}
\mbox{\epsfig{file=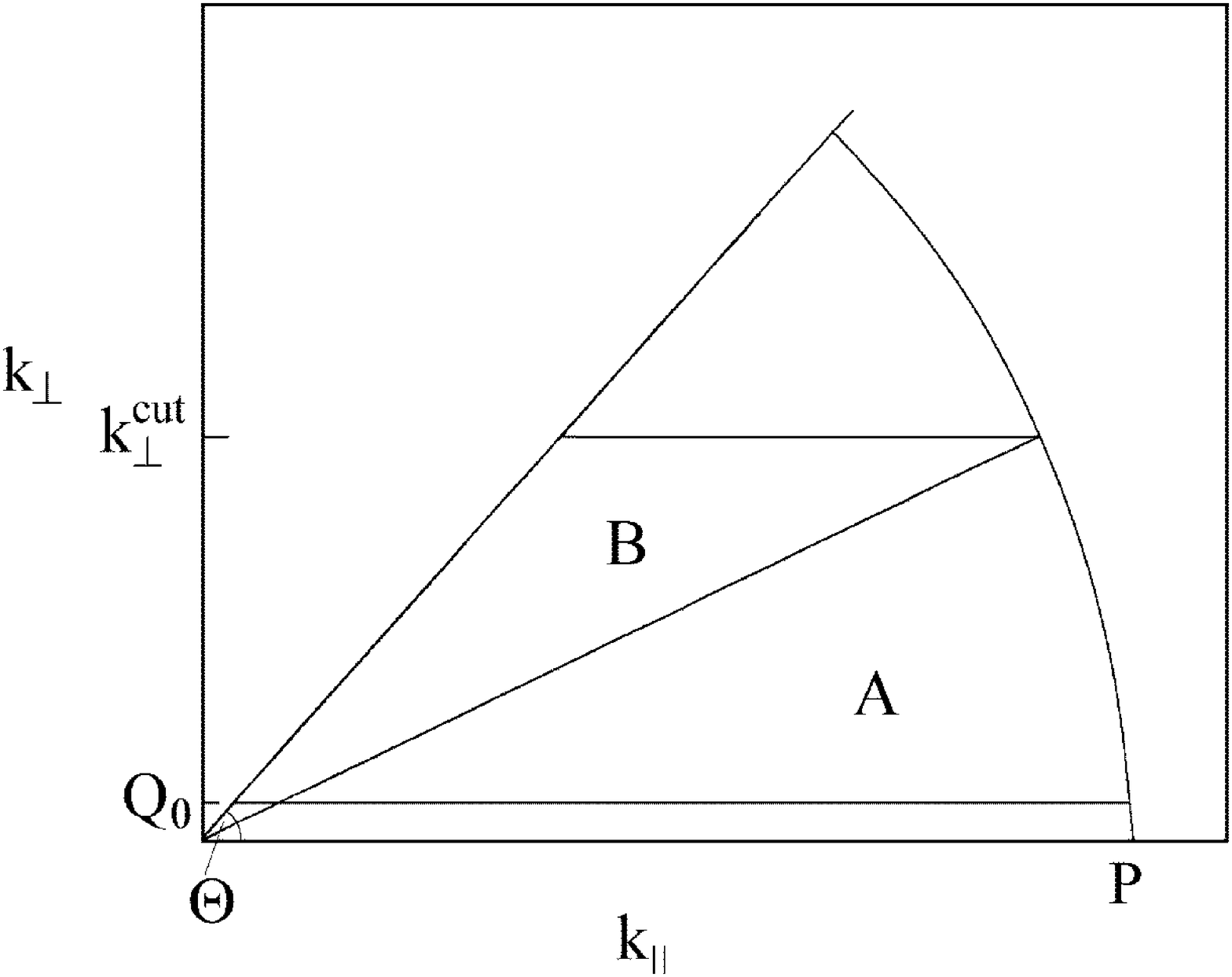,bbllx=-10.cm,bblly=4.cm,bburx=25.cm,bbury=28.cm,width=10cm}}
          \end{center}
\caption{}
\label{kin3}
\end{figure}

\noindent
{\bf~Fig.~\ref{kin3}}
Kinematics relevant for the splitting the cylindrically cut moments,
c.f. Eqs.(\eref{nsplit},\eref{c2split}). Partons in the region A
form a complete jet with virtuality $\kct$.

\begin{figure}[p]
\vspace{-2.0cm} 
 \begin{minipage}{.95\linewidth}
          \begin{center}
\mbox{\mbox{\epsfig{file=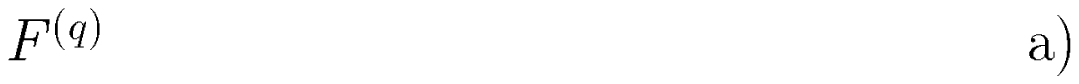,bbllx=5.cm,bblly=20.cm,bburx=5.2cm,bbury=24.cm,width=0.2cm}}
\mbox{\epsfig{file=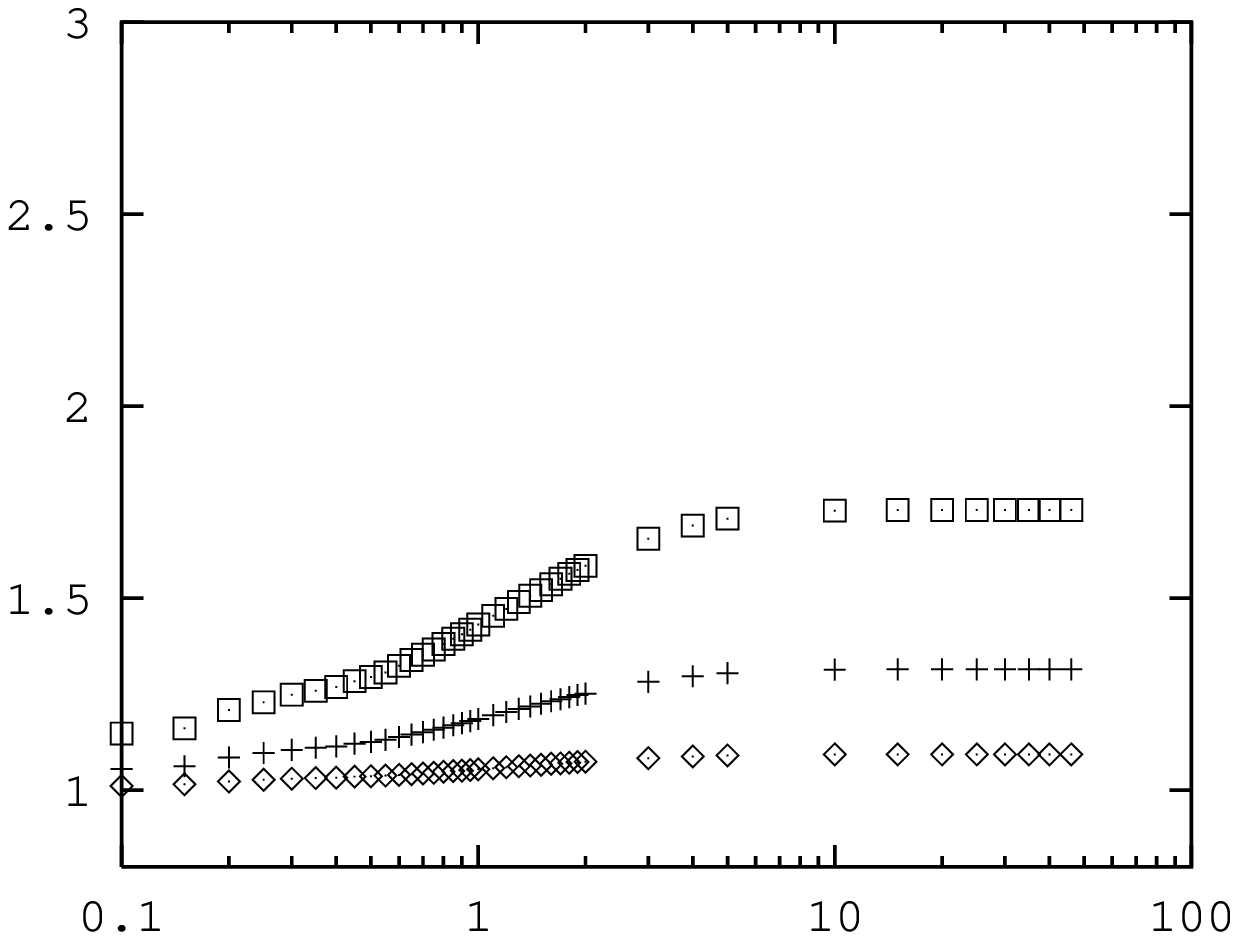,bbllx=3.cm,bblly=2.5cm,bburx=17.cm,bbury=14.5cm,width=12.cm}}
      }         \end{center}
\vspace{-1.0cm}
\hspace{6.2cm} \large $k_\perp^{\mathrm cut}$ [GeV]
      \end{minipage}
      \begin{minipage}{.95\linewidth}
          \begin{center}
\mbox{\mbox{\epsfig{file=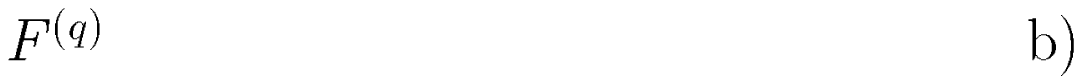,bbllx=5.cm,bblly=20.cm,bburx=5.2cm,bbury=24.cm,width=0.2cm}}
\mbox{\epsfig{file=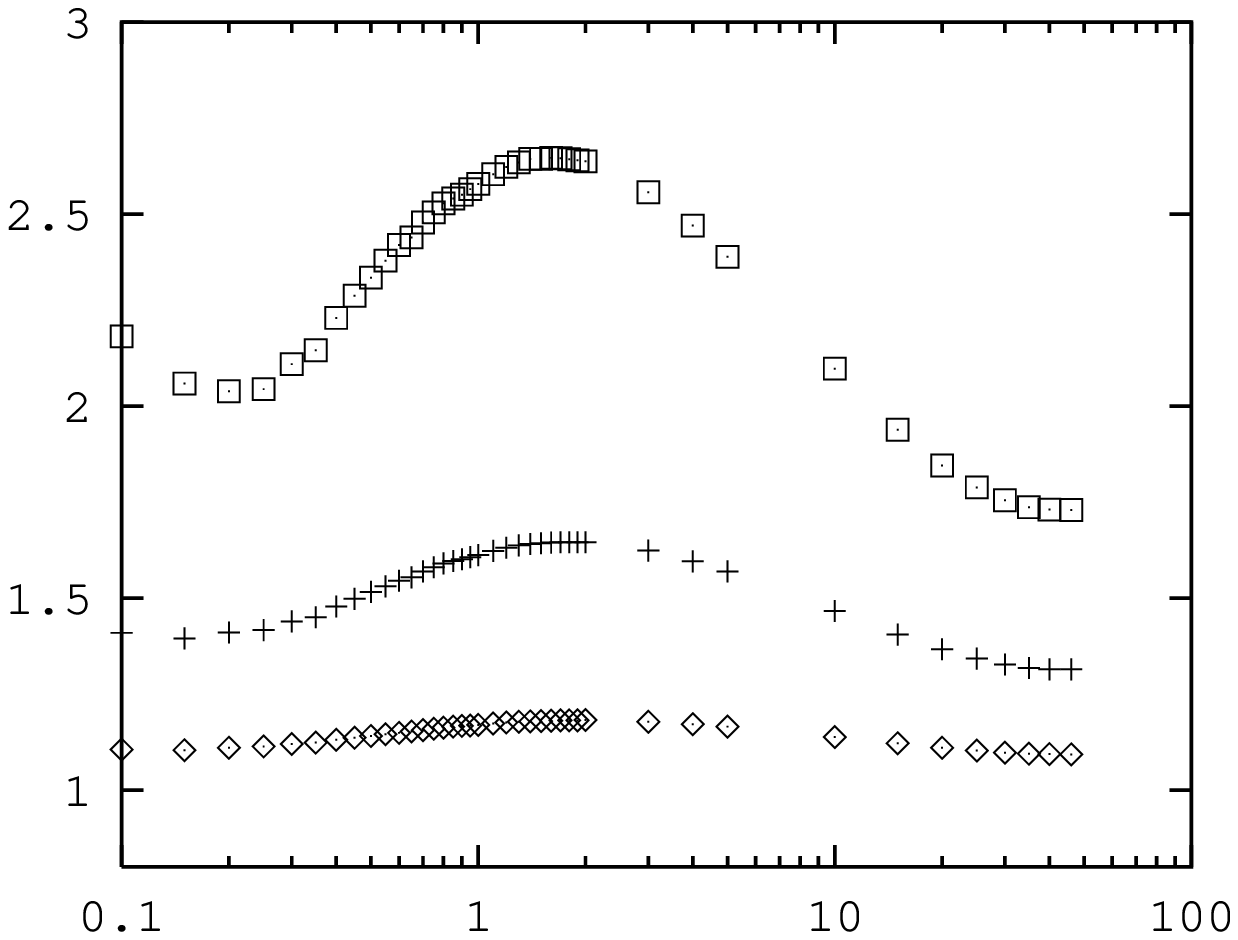,bbllx=3.cm,bblly=2.5cm,bburx=17.cm,bbury=14.5cm,width=12.cm}}
      }           \end{center}
\vspace{-1.0cm}
\hspace{6.2cm} \large $k^{\mathrm cut}$ [GeV]
      \end{minipage}
\caption{}
\label{figpart02}
\end{figure}

\medskip
\noindent 
{\bf Fig.~\ref{figpart02}}. {\bf a}: cut moments of order 2 (diamonds), 
 (crosses) and 4 (squares) in one hemisphere defined through the thrust
axis as a function of $k_\perp^{cut}$ as predicted 
by ARIADNE at parton level 
with parameters $Q_0$ = 0.2 GeV and $\lambda$ = 0.015 
at $\sqrt{s}$ = 91.2 GeV. {\bf b}: same as in {\bf a}, but as a function of 
$k^{cut}$. 

\begin{figure}[p]
\vspace{-2.0cm} 
 \begin{minipage}{.95\linewidth}
          \begin{center}
\mbox{\mbox{\epsfig{file=label.ps,bbllx=5.cm,bblly=20.cm,bburx=5.2cm,bbury=24.cm,width=0.2cm}}
\mbox{\epsfig{file=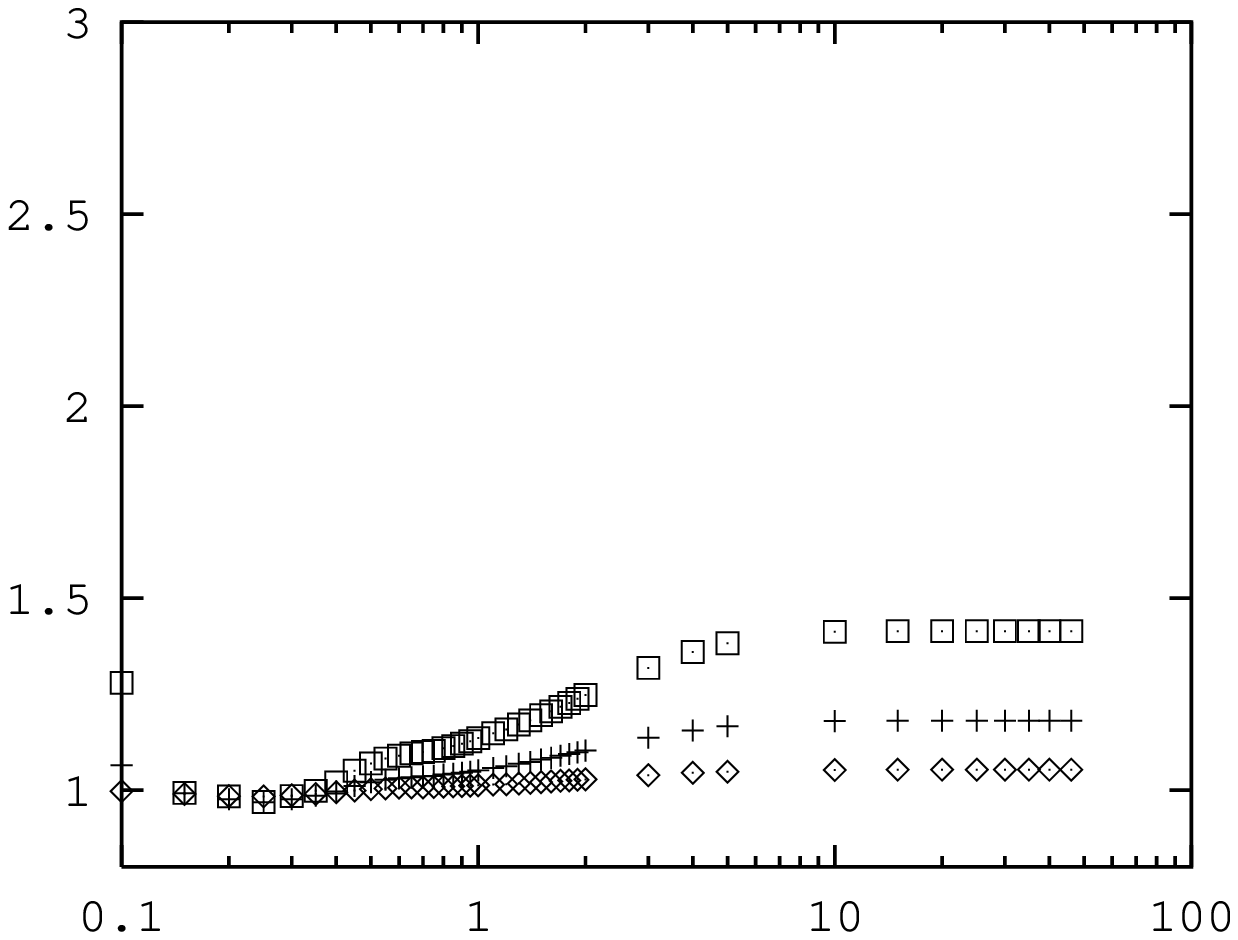,bbllx=3.cm,bblly=2.5cm,bburx=17.cm,bbury=14.5cm,width=12.cm}}
       }          \end{center}
\vspace{-1.0cm}
\hspace{6.2cm} \large $k_\perp^{\mathrm cut}$ [GeV]
      \end{minipage}
      \begin{minipage}{.95\linewidth}
          \begin{center}
\mbox{\mbox{\epsfig{file=labelb.ps,bbllx=5.cm,bblly=20.cm,bburx=5.2cm,bbury=24.cm,width=0.2cm}}
\mbox{\epsfig{file=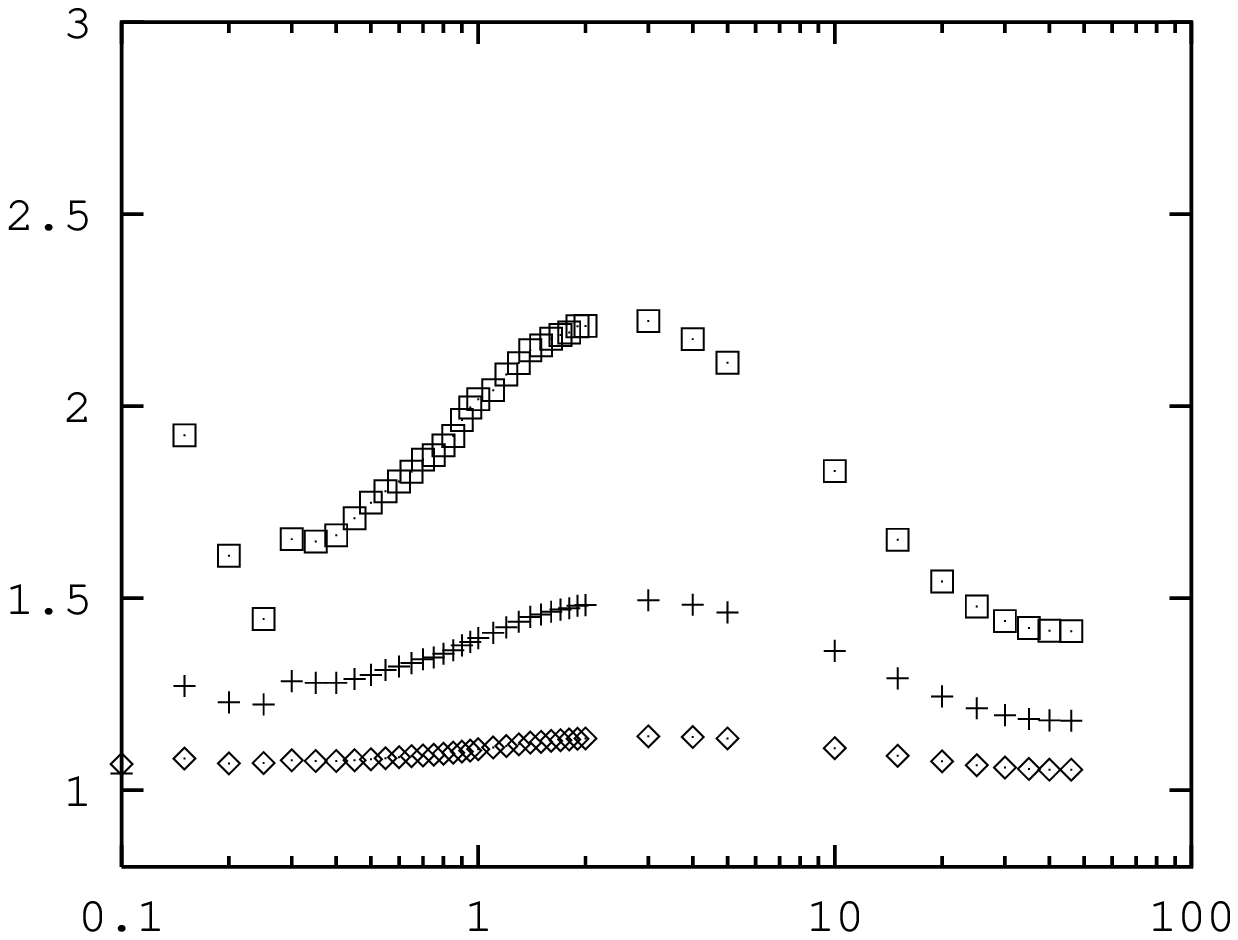,bbllx=3.cm,bblly=2.5cm,bburx=17.cm,bbury=14.5cm,width=12.cm}}
      }            \end{center}
\vspace{-1.0cm} 
\hspace{6.2cm} \large $k^{\mathrm cut}$ [GeV]
      \end{minipage}
\caption{}
\label{figpart}
\end{figure}

\medskip 
\noindent 
{\bf Fig.~\ref{figpart}}. Same as in Fig.~{\bf \ref{figpart02}}, but with $Q_0$ = 0.4 GeV. 

\begin{figure}[p]
\vspace{-2.0cm}
 \begin{minipage}{.95\linewidth}
          \begin{center}
\mbox{\mbox{\epsfig{file=label.ps,bbllx=5.cm,bblly=20.cm,bburx=5.2cm,bbury=24.cm,width=0.2cm}}
\mbox{\epsfig{file=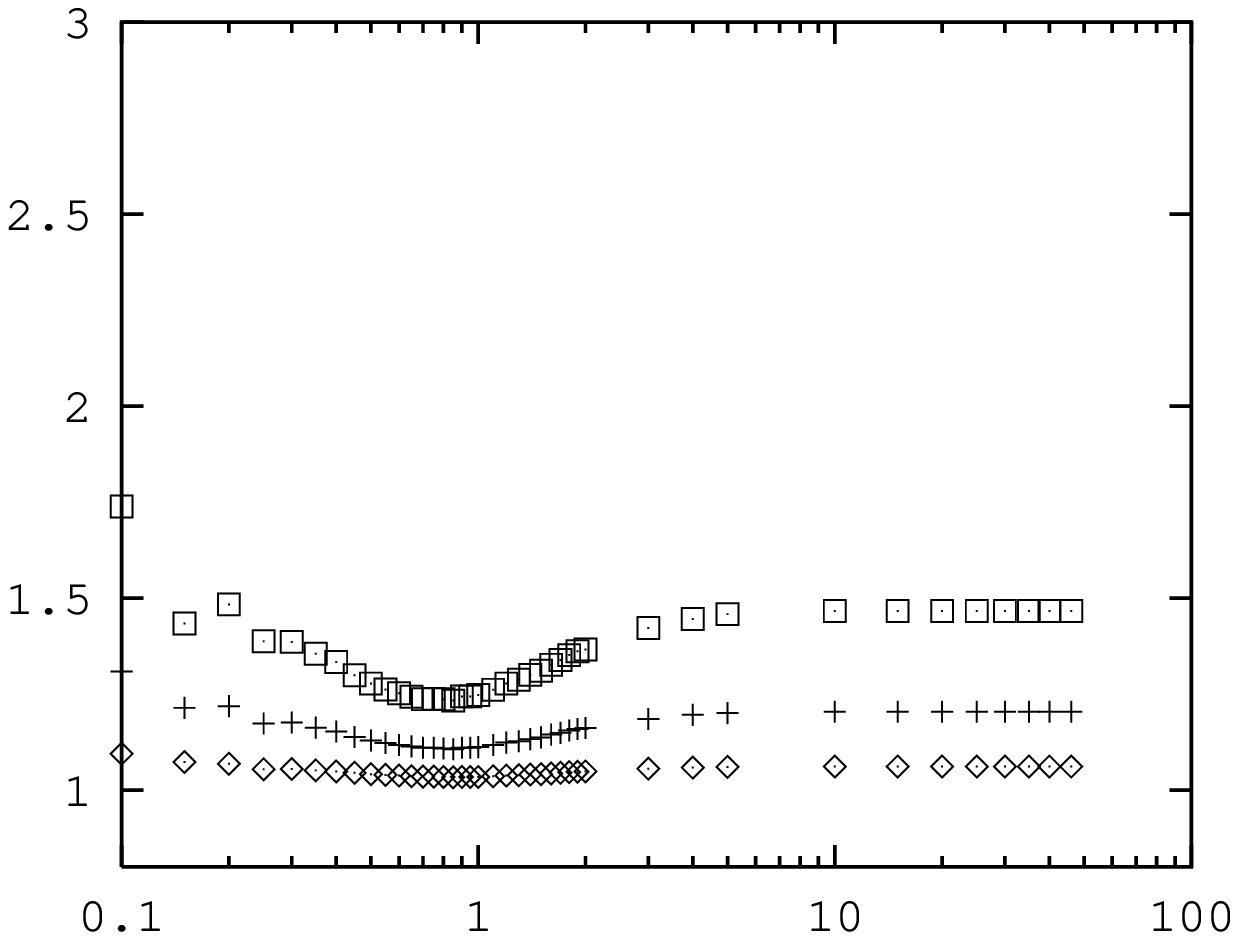,bbllx=3.cm,bblly=2.5cm,bburx=17.cm,bbury=14.5cm,width=12cm}}
      }           \end{center}
\vspace{-1.0cm}
\hspace{6.2cm} \large $k_\perp^{\mathrm cut}$ [GeV]
      \end{minipage}
      \begin{minipage}{.95\linewidth}
          \begin{center}
\mbox{\mbox{\epsfig{file=labelb.ps,bbllx=5.cm,bblly=20.cm,bburx=5.2cm,bbury=24.cm,width=0.2cm}}
\mbox{\epsfig{file=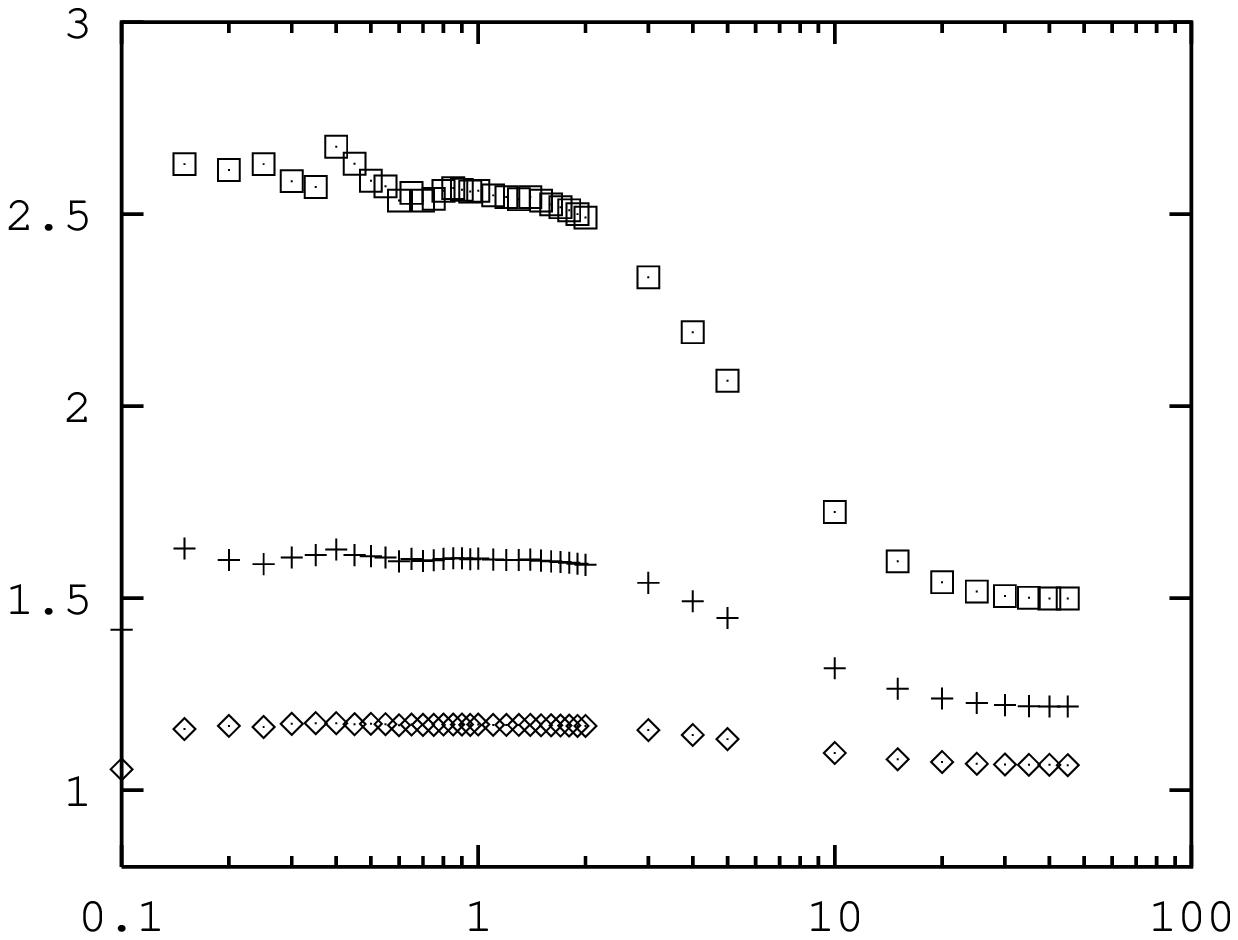,bbllx=3.cm,bblly=2.5cm,bburx=17.cm,bbury=14.5cm,width=12cm}}
      }           \end{center}
\vspace{-1.0cm}
\hspace{6.2cm} \large $k^{\mathrm cut}$ [GeV]
      \end{minipage}
\caption{}
\label{fighadr}
\end{figure}

\medskip 
\noindent 
{\bf Fig.~\ref{fighadr}}. Same as in Fig.~{\bf \ref{figpart02}}, but at hadron level with default 
values of the parameters and string fragmentation. 

\newpage

\newpage
 
\newpage

\end{document}